\documentclass[a4paper,11pt,openany]{article}
\usepackage{pst-3d}
\usepackage{pst-3dplot}
\usepackage{etex}
\usepackage[utf8]{inputenc}
\usepackage[english]{babel}
\usepackage{amsmath,amsfonts,amssymb,amsthm,mathabx}
\usepackage{graphicx}
\usepackage[footnotesize, skip=0pt]{caption}
\usepackage{epigraph}
\usepackage{indentfirst}
\usepackage{booktabs}
\usepackage{fancyhdr}
\usepackage{vmargin}
\usepackage{feynmf}
\usepackage{yfonts}
\usepackage{lscape}
\usepackage{subfig}
\usepackage{textcomp}
\usepackage{pstricks}
\usepackage[metapost]{mfpic}
\usepackage{fancybox}
\usepackage{slashed}
\usepackage[verbose]{wrapfig}
\usepackage{pifont}
\usepackage{keystroke}
\usepackage{appendix}
\usepackage{enumitem}
\usepackage{array}
\usepackage{colortbl}
\usepackage{alltt}
\usepackage{boxedminipage}
\usepackage{tikz}
\usepackage{calc}
\usepackage{subfloat}
\usepackage{multirow}
\usepackage{cmll}
\usepackage{multicol}
\definecolor{color1}{RGB}{204,0,51}
\definecolor{color2}{RGB}{159,182,205}
\usepackage{bookmark}

\usepackage{verbatim}
\usepackage[vcentermath]{youngtab}
\usepackage{young}
\usepackage{pst-grad} 
\usepackage{pst-plot}
\usepackage{transparent}
\usepackage{color,soul}
\usepackage{cancel}
\usepackage{bm}
\usepackage{pifont}
\usepackage{placeins}
\usepackage{mathtools}
\usepackage{rotating}
\usepackage[refpage]{nomencl}
\usepackage{hhline}
\usepackage{marginnote}
\usepackage{upgreek}
\usepackage{stackengine}
\usepackage{setspace}
\usepackage{ytableau}
\usepackage{listings}
\usepackage{calligra}
\usepackage{hyperref}
\usepackage{afterpage}
\usepackage{cleveref}
\usepackage{makecell}
\usepackage{accents}
\usepackage{breqn}
\usepackage{environ}
\usepackage{cite}

\numberwithin{equation}{section}

\def\beq{\begin{equation}}
\def\eeq{\end{equation}}
\def\bay{\begin{array}}
\def\eay{\end{array}}

\allowdisplaybreaks

\sethlcolor{yellow}
\definecolor{boh}{RGB}{79,47,79}

\hypersetup{
linkbordercolor={red}
}

\hypersetup{
    bookmarks=false,         
    unicode=false,          
    pdftoolbar=true,        
    pdfmenubar=true,        
    pdffitwindow=false,     
    pdfstartview={FitH},    
    pdftitle={Duality between Galilei and Carroll from Brane Point of View},    
    pdfauthor={Eric Bergshoeff, Luca Romano},     
    pdfsubject={},   
    pdfnewwindow=true,      
    colorlinks=false,       
    linkcolor=red,          
    citecolor=red,        
    filecolor=red,      
    urlcolor=red           
    linkbordercolor={red},
    citebordercolor={red},
    urlbordercolor={red}
}

\makeatletter

\newcommand{\Rmnum}[1]{\expandafter\@slowromancap\romannumeral #1@}
\makeatother

\theoremstyle{definition}

\theoremstyle{remark}

\theoremstyle{proposition}

\usepackage{alphalph}

\makeatletter
\newalphalph{\aalphalph}[mult]{\alphalph@alph}{26}
\newcommand{\alphalphval}[1]{%
  \@ifundefined{c@#1}{
    \aalphalph{#1}
  }{%
    \aalphalph{\value{#1}}
  }
}
\makeatother

\AtBeginEnvironment{subequations}{%
  \let\alph\alphalphval%
}

\usepackage{color}

\def\chapterautorefname~#1\null{Chap.~(#1)\null}
\def\sectionautorefname~#1\null{Sec.~(#1)\null}
\def\subsectionautorefname~#1\null{sub--Sec.~(#1)\null}
\def\figureautorefname~#1\null{Fig.~(#1)\null}
\def\tableautorefname~#1\null{Tab.~(#1)\null}
\def\equationautorefname~#1\null{eq.~(#1)\null}

\def\equationautorefname~#1\null{eq.~(#1)\null}

\NewEnviron{multieq}[1][2]{

\begin{multicols}{#1}
\begin{subequations}
\setlength{\abovedisplayskip}{-12pt}
\allowdisplaybreaks
\begin{align}
\BODY
\end{align}
\end{subequations}
\end{multicols}
\setlength{\parindent}{0pt}
}

\NewEnviron{multieqsep}[1][2]{

\begin{multicols}{#1}
\setlength{\columnseprule}{0.4pt}
\begin{subequations}
\setlength{\abovedisplayskip}{-12pt}
\allowdisplaybreaks
\begin{align}
\BODY
\end{align}
\end{subequations}
\end{multicols}
\setlength{\parindent}{0pt}
}

\DeclareMathAlphabet\mathbfcal{OMS}{cmsy}{b}{n}

\title{\bf Non-relativistic limits and three-dimensional coadjoint Poincar\'e gravity}

\begin{document}

\begin{flushright}
\small
ICCUB-19-019\\
\normalsize
\end{flushright}
{\let\newpage\relax\maketitle}
\maketitle
\def\equationautorefname~#1\null{eq.~(#1)\null}
\def\tableautorefname~#1\null{tab.~(#1)\null}

\vspace{0.8cm}

\begin{center}

\renewcommand{\thefootnote}{\alph{footnote}}
{\sl\large Eric Bergshoeff$^{~1}$}\footnote{Email: {\tt e.a.bergshoeff[at]rug.nl}},
{\sl\large Joaquim Gomis$^{~2}$}\footnote{Email: {\tt joaquim.gomis[at]ub.edu}} and
{\sl\large Patricio Salgado-Rebolledo$^{~3}$}\footnote{Email: {\tt patricio.salgado[at]pucv.cl}}

\setcounter{footnote}{0}
\renewcommand{\thefootnote}{\arabic{footnote}}

\vspace{0.5cm}

${}^1${\it Van Swinderen Institute, University of Groningen\\
Nijenborgh 4, 9747 AG Groningen, The Netherlands}\\
\vskip .2truecm
${}^2${\it Departament de F\'{\i}sica Cu\`antica i Astrof\'{\i}sica and Institut de Ci\`encies del\\
Cosmos, Universitat de Barcelona, Mart\'i i Franqu\`es 1, E-08028 Barcelona, Spain}\\
\vskip .2truecm
${}^3${\it Instituto de F\'isica, Pontificia Universidad Cat\'olica de Valpara\'iso\\ Casilla 4059, Valparaiso, Chile}\\

\vspace{1.8cm}


{\bf Abstract}
\end{center}
\begin{quotation}
  {\small
We show that a  recently proposed action  for three-dimensional non-relativistic gravity can be obtained by taking the limit of a relativistic Lagrangian that involves the co-adjoint Poincar\'e algebra. We point out the similarity of our construction with the way that three-dimensional Galilei Gravity and Extended Bargmann Gravity can be obtained by taking the limit of a relativistic Lagrangian that involves the Poincar\'e algebra. We extend our results to the AdS case  and we will see that there is a chiral decomposition both at the relativistic and non-relativistic level. We comment on possible further generalizations}.
\end{quotation}

\newpage
\tableofcontents

\newpage
\section{Introduction}
Lie algebra contractions are a useful technique to obtain non-relativistic (NR) symmetries from relativistic ones \cite{Inonu:1953sp}. Within this context, the Galilei algebra can be understood as an In\"on\"u-Wigner contraction of the Poincar\'e algebra when the speed of light goes to infinity. This procedure can be generalized to obtain the Bargmann algebra \cite{LL,BGGK}, which allows one to properly implement the NR limit of a relativistic particle at the level of the action \footnote{Note that when considering a $(p+1)$-dimensional extended object there are $p+1$ possible NR limits \cite{Batlle:2016iel,Barducci:2018wuj}.}.


The In\"on\"u-Wigner contraction can be generalized to increase the number of Lie algebra generators by considering Lie algebra expansions \cite{Hatsuda:2001pp,deAzcarraga:2002xi,Izaurieta:2006zz}. Indeed, NR expansions of the Poincar\'e algebra lead to an infinite family of extensions of the Galilei algebra~\cite{Bergshoeff:2019ctr}, which have been shown to underlie the large c expansion of General Relativity~\cite{Hansen:2018ofj,Hansen:2019vqf}. This method has been extended to the case of strings \cite{Bergshoeff:2019ctr,Harmark:2019upf} and the case of non-vanishing cosmological constant \cite{Gomis:2019nih}.
Another way to obtain such a sequence of NR symmetries is by considering suitable quotients of a Galilean free Lie algebra construction~\cite{Gomis:2019fdh}.

Concerning applications of these algebras to physical systems, one should  distinguish between a) limits of sigma model actions for particles, strings or
p-branes; and b) limits of target space actions leading to NR gravity. In the first case, it is only known how to define NR limits for the Galilei and Bargmann algebra, but not for further extensions of these symmetries. As far as gravitational actions are concerned, it is possible to construct gravitational actions invariant under extensions of the Bargmann algebra using the Lie algebra expansion method. However, to obtain these actions as NR limits of relativistic action is more subtle due to the possible appearance of infinities. For
example, the formulation of a NR limit of the 4D Einstein-Hilbert action leading to a finite action of Newton-Cartan gravity is a well-known open problem,
see for example \cite{DePietri:1994je,Bergshoeff:2017btm} and references therein\footnote{For a recent proposal for an action for Newtonian gravity, see \cite{Hansen:2018ofj}. Note that the limit exist in the case of four-dimensional String Newton-Cartan gravity, but only after a term has been added to the Einstein-Hilbert action \cite{Bergshoeff:2018vfn}. In section 4 we will give a new example of a finite limit of the EH action without the need to add an extra term.}.

In the case of three dimensions, where gravity can be formulated as a Chern-Simons (CS) gauge theory \cite{Achucarro:1987vz,Witten:1988hc},
the limit
can be studied in a more transparent way. In reference \cite{Bergshoeff:2016lwr} it was shown that a consistent NR limit of three-dimensional Einstein gravity involving an extra CS term with two $\mathfrak{u}(1)$ gauge fields
is given by a NR CS action
invariant under the Extended Bargmann algebra \cite{Papageorgiou:2009zc}. The inclusion of a cosmological constant in this NR gravity theory has been studied in detail  in \cite{Papageorgiou:2010ud,Hartong:2016yrf}.

It is natural to address the question whether the NR symmetries and corresponding gravity actions of \cite{Hansen:2018ofj,Hansen:2019vqf,Bergshoeff:2019ctr,Ozdemir:2019orp}
can be obtained as the NR limit of an enlarged Poincar\'e symmetry algebra and a corresponding gravity theory, respectively. Concerning the algebra, a natural possible candidate is the so-called coadjoint Poincar\'e symmetry. This algebra and some of its contractions have been studied in
\cite{Barducci:2019jhj} to obtain the p-brane Galilei algebra
\cite{Brugues:2004an,Gomis:2005pg,Brugues:2006yd,Barducci:2018wuj,Gomis:2019fdh}. These contractions could be useful to obtain the NR string theories of
\cite{Gomis:2000bd,Danielsson:2000gi,Bergshoeff:2018vfn,Gomis:2019zyu} as the  limit of a relativistic string theory with an enlarged relativistic space-time symmetry algebra. In this paper we will show that the coadjoint  Poincar\'e algebra indeed provides a relativistic counterpart of the NR algebra introduced in \cite{Hansen:2018ofj}.

On the other hand, the algebra found in \cite{Ozdemir:2019orp} can be obtained from the coadjoint Poincar\'e algebra plus two $\mathfrak{u}(1)$ generators, leading to two central extensions. We will refer to this algebra as the Enhanced Bargmann algebra.  Furthermore, in \cite{Ozdemir:2019orp}  it was shown that the corresponding gravity action can be obtained as a limit of a CS action based on the $\mathfrak{iso}(2,1)\oplus \mathbb{E}_3\oplus \mathfrak{u}(1)^2$ algebra. In this paper we show that the same action can be obtained from a fully relativistic CS theory invariant under the direct sum of the coadjoint Poincar\'e algebra and two $\mathfrak{u}(1)$ factors. Furthermore, it is possible to obtain Extended Bargmann gravity and Galilei gravity as alternative NR limits of the same action without generating infinities.

We point out the similarity of our construction with the way in which (2+1)-dimensional Galilei Gravity \cite{Bergshoeff:2017btm} and Extended Bargmann Gravity \cite{Papageorgiou:2009zc,Bergshoeff:2016lwr,Hartong:2016yrf} can be obtained as a limit of a relativistic CS action invariant under the Poincar\'e algebra. The results here obtained can be generalized in several ways. First, we discuss the extension to the case of the coadjoint AdS algebra by including a cosmological constant and show that it is possible to define a chiral decomposition both at the relativistic and NR level. This decomposition is the analogue of the $\mathfrak sl(2\mathbb R)$ formulation of AdS$_3$ CS gravity \cite{Achucarro:1987vz,Witten:1988hc} or the chiral decomposition of AdS$_3$ invariant dynamical systems and their NR counterparts \cite{Alvarez:2007ys,Alvarez:2007fw,Batlle:2014sca}.
Second, in the outlook we argue that the coadjoint Poincar\'e algebra can be defined as a particular relativistic expansion of the Poincar\'e symmetry. Based of this fact, we suggest  a generalization of our construction to extensions of the coadjoint Poincar\'e algebra that result from using bigger semigroups \cite{Izaurieta:2006zz} in the expansion of the Poincar\'e algebra.


The organization of the paper is as follows: in Section \ref{NRP}, we consider contractions of the Poincar\'e algebra and the enlarged Poincar\'e $\oplus\ \mathfrak{u}(1)^2$ algebra. We consider the corresponding NR limit of the CS actions that are based upon these two algebras and show that they give rise to Galilei and Extended Bargmann gravity. In Section \ref{NRCP}, we repeat the same analysis but now for the  coadjoint Poincar\'e and the enlarged coadjoint Poincar\'e $\oplus\ \mathfrak{u}(1)^2$ algebras. In particular, we show that this time the CS actions lead to the actions of not only Galilei and Extended Bargmann gravity but also to the action of Enhanced Bargmann gravity \cite{Ozdemir:2019orp}. In Section \ref{NRCAdS}, we discuss the coadjoint AdS algebra and generalize our results to include a cosmological constant. In the Conclusions we speculate how the results obtained in this paper can be generalized to construct gravity actions based on further extensions of the Enhanced Bargmann algebra.

\section{NR limits and the Poincar\'e algebra}
\label{NRP}

In this section, we consider how the action for three-dimensional Galilei gravity \cite{Bergshoeff:2017btm} and the action for extended Bargmann gravity  \cite{Papageorgiou:2009zc,Bergshoeff:2016lwr,Hartong:2016yrf} can be obtained by taking the limit of specific relativistic actions.
This section serves as a warming up exercise for the next section where we will go beyond Extended Bargmann gravity and reproduce the Enhanced Bargmann gravity action of \cite{Ozdemir:2019orp}
as the limit of a relativistic CS action invariant under the co-adjoint Poincar\'e algebra following the same construction that we perform in this section.

\subsection{The Poincar\'e algebra}

Our starting point is the $D$-dimensional Poincar\'e algebra of space-time translations $\tilde P_{ A}$ and Lorentz transformations $\tilde J_{ A B}\ (A=0,1,\dots, D-1)$
\begin{equation}\label{Poincare}
[\tilde J_{ A B},  \tilde P_{ C}] =  2\eta_{ C[ B}\tilde P_{ A]}\,,\hskip 2.5 truecm
[\tilde J_{ A B}, \tilde J_{ C D}] = 4\eta_{[ A[ C}\tilde J_{ D] B]}\,,
\end{equation}
where $\eta_{ A B}$ is the (mostly plus) Minkowski metric. Splitting the relativistic indices in temporal and spatial values as $A=(0,a)$, where $a=1,\dots,D-1$, we can write
\begin{equation}\label{splittingJP}
\tilde J_{AB} =\big\{\tilde J_{0a}\equiv \tilde G_a, \tilde J_{ab}\big\}\,,\hskip 2.5 truecm
\tilde P_A = \big\{\tilde P_0\equiv \tilde H, \tilde P_a \big\}\,,
\end{equation}
and the commutation relations \eqref{Poincare} take the form
\begin{multicols}{2}
\begin{subequations} \label{splitpoincare}
\setlength{\abovedisplayskip}{-14pt}
\allowdisplaybreaks
\begin{align}
[\tilde G_{a},\tilde H]&=\tilde P_{a}\, ,\\[.1truecm]
[\tilde G_{a},\tilde P_{b}]&=\delta_{ab}\tilde H\, , \\[.1truecm]
[\tilde G_{a},\tilde G_{b}]&=\tilde J_{ab}\,,\\[.1truecm]
[\tilde J_{ab},\tilde J_{cd}]&=4\delta_{[a[c}\tilde J_{d]b]}\, ,\\[.1truecm]
%
%
%
[\tilde J_{ab},\tilde G_{c}]&= 2\delta_{c[b}\tilde G_{a]}\,,\\[.1truecm]
[\tilde J_{ab},\tilde P_{c}]&= 2\delta_{c[b}\tilde P_{a]}\,.
\end{align}
\end{subequations}
\end{multicols}\noindent
From these relations we see that the Galilei algebra can be obtained by means of the following rescaling of the Poincar\'e generators\footnote{\label{footnote1} This scaling of the generators is different from  the scaling used in \cite{Gomis:2019sqv} and resembles more the  ones of \cite{Barducci:2019fjc}.  Other scalings are also possible
\cite{Papageorgiou:2009zc,Bergshoeff:2016lwr,Hartong:2016yrf}. This is related to the fact that the commutation relations of the Poincar\'e algebra are invariant under the rescaling $\tilde H =\lambda H$ and $\tilde P_a = \lambda \tilde P_a$
(see for example \cite{Barducci:2019fjc}). At the field theory level it corresponds to the fact that two different scalings can differ by an overall scaling of the Lagrangian that can be absorbed by a scaling of Newton's constant
\cite{Bergshoeff:2019ctr}.}

\begin{multicols}{2}
\begin{subequations}\label{redefinitionP}
\setlength{\abovedisplayskip}{-14pt}
\allowdisplaybreaks
\begin{align}
 J_{ab} &= \tilde J_{ab}\,,\label{Jabredef} \\[.1truecm]
 H &= \tilde H\,, \label{Hredef}  \\[.1truecm]
 \tilde G_a &= \varepsilon  G_a\,, \label{Gredef}\\[.1truecm]
\tilde P_a &= \varepsilon  P_a\,,\label{Predef}
\end{align}
\end{subequations}
\end{multicols}\noindent
and taking the limit $\varepsilon\rightarrow \infty$, which leads to
\begin{multicols}{2}
\begin{subequations} \label{GalileiD}
\setlength{\abovedisplayskip}{-14pt}
\allowdisplaybreaks
\begin{align}
[ G_{a},H]&= P_{a}\, ,\\[.1truecm]
[ J_{ab}, J_{cd}]&=4\delta_{[a[c} J_{d]b]}\, ,\\[.1truecm]
%
%
%
[ J_{ab}, G_{c}]&= 2\delta_{c[b} G_{a]}\,,\\[.1truecm]
[ J_{ab}, P_{c}]&= 2\delta_{c[b} P_{a]}\,.
\end{align}
\end{subequations}
\end{multicols}\noindent

In $D=2+1$ dimensions, we can rewrite the Poincar\'e commutation relations \eqref{Poincare} by  defining the dual generators
\begin{equation}\label{dualJ}
\tilde J_{ A B} \equiv \epsilon^C_{\;\;AB} \tilde J_C
\end{equation}
in the following way:
\begin{equation}\label{dualP}
[\tilde J_A, \tilde J_B] = \epsilon^C_{\;\;AB}\tilde J_C\,,\hskip 2truecm [\tilde J_A, \tilde P_B] = \epsilon^C_{\;\;AB}\tilde P_C\,,
\end{equation}
where we have defined  the epsilon-tensor such that $\epsilon_{012} = -1$. Using the dual generator \eqref{dualJ} requires to replace the relation for $\tilde J_{AB}$ in \eqref{splittingJP} by
\begin{equation}\label{splitJA}
J_A=\big\{\tilde J_{a}\equiv \tilde G_a\,,\; \tilde J_{0}\equiv \tilde J\big\}\,,
\end{equation}
and to replace
\begin{equation}\label{dualNRJG}
J_{ab}=-\epsilon_{ab}J\,,\hskip 2truecm G_a\rightarrow -\epsilon_a^{\;\;b}G_a
\end{equation}
in the contraction \eqref{redefinitionP}. This leads to the following form for the Galilei algebra in three space-time dimensions
\begin{multicols}{2}
\begin{subequations}\label{Galilei}
\setlength{\abovedisplayskip}{-14pt}
\allowdisplaybreaks
\begin{align}
[J, G_a] &= -\epsilon_a{}^b G_b\,,\\[.1truecm]
[J, P_a] &= -\epsilon_a{}^b P_b\,,\\[.1truecm]
[H, G_a] &= -\epsilon_a{}^bP_b\,,
\end{align}
\end{subequations}
\end{multicols}\noindent
where the two-dimensional epsilon-tensor is defined such that $\epsilon_{ab}\equiv-\epsilon_{0ab}\Rightarrow\epsilon_{12}=1$.

In 2+1 dimensions we can define a gravity theory invariant under the full Poincar\'e algebra by considering the CS action \cite{Achucarro:1987vz,Witten:1988hc}:
\begin{equation}\label{CSaction}
S=  \int \left\langle A\wedge dA + \frac{2}{3}A\wedge A\wedge A\right\rangle \,,
\end{equation}
where the gauge field $A$  takes values in the Poincar\'e algebra:
\begin{equation}
A = \Omega^A\tilde J_A + E^A\tilde P_A\,.
\end{equation}
We consider a non-degenerate invariant bilinear  form of the Poincar\'e algebra:
\begin{equation}\label{pairingR}
 \left\langle \tilde J_A \tilde P_B \right\rangle = \alpha_1\eta_{AB}\,,
\end{equation}
with arbitrary parameter $\alpha_1\ne 0$.

The curvature two-form of the Poincar\'e algebra reads
\begin{equation}
R = dA + A^2 = R^A(\tilde J)\tilde J_A + R^A(\tilde P)\tilde P_A\,,
\end{equation}
where
\begin{equation}\label{curvaturecompP}
R^A(\tilde J) = d\Omega^A +\frac{1}{2}\epsilon^A_{\;\;BC}
\Omega^B\Omega^C\,, \hskip .5truecm
R^A(\tilde P) = d E^A + \epsilon^A_{\;\;BC}\Omega^B E^C\,.
\end{equation}
Explicitly, up to boundary terms, one finds the three-dimensional version of the Einstein-Hilbert term as the CS action action for Poincar\'e gravity, i.e.
\begin{equation}\label{explicit}
S_{\textrm{Poincar\'e}} = 2\alpha_1 \int E^A R^B(\tilde J)\eta_{AB}\,.
\end{equation}
Using \eqref{redefinitionP} and \eqref{dualNRJG}, the connection $A$ can be expressed as
\begin{equation}
A = \omega J + \omega^a G_a +\tau H + e^a P_a\,,
\end{equation}
where the NR gauge fields are related to the relativistic ones by
\begin{multicols}{2}
\begin{subequations}\label{gfcontractionP}
\setlength{\abovedisplayskip}{-13pt}
\allowdisplaybreaks
\begin{align}
\Omega^0&=\omega\,,\\[.1truecm]
\Omega^a &= \frac{1}{\varepsilon}\omega^a\,,\\[.1truecm]
E^0 &=\tau\,,\\[.1truecm]
E^a &= \frac{1}{\varepsilon} e^a\,.
\end{align}
\end{subequations}
\end{multicols}
\noindent
Using these relations, the Poincar\'e gravity  action \eqref{explicit} takes the form
\begin{equation}\label{Poincaregrav}
S_{\textrm{Poincar\'e}} = 2\alpha_1\int \left[-\tau R(J) +\frac{1}{\varepsilon^2} e^a R_a(G)\right]\,,
\end{equation}
where
\begin{equation}\label{RJRG}
R(J)=d\omega\,,\hskip 1.5truecm
R^a(G)=d\omega^a -\epsilon^{a}_{\;\;b}\,\omega\,\omega^b
\end{equation}
are the components of the NR curvature two-form of the Galilei algebra corresponding to spatial rotations and Galilean boosts, respectively.
\vskip .3truecm
\noindent {\bf Galilei gravity}
\vskip .1truecm

In the limit that $\varepsilon\rightarrow \infty$, the action for Poincar\'e gravity reduces to the action for Galilei gravity
\cite{Bergshoeff:2017btm}
\begin{equation}\label{Gaction}
S_{\textrm{Galilei}} =  -2\kappa \int \tau R(J) \,,
\end{equation}
were we have set $\alpha_1=\kappa$, which plays the role of the gravitational coupling constant. We note that in this limit the invariant tensor \eqref{pairingR} gives the NR invariant bilinear form\,\footnote{
We note that the most general invariant tensor for the Galilei algebra is degenerate and is given by
\begin{equation*}
\left\langle J,J \right\rangle =-\beta_1\,,\quad
\left\langle J,H \right\rangle =-\beta_2\,,\quad
\left\langle H,H \right\rangle =-\beta_3\,,
\end{equation*}
with $\beta_1, \beta_2, \beta_3$ arbitrary parameters. The first term comes from the relativistic invariant form $\left\langle \tilde J_A \tilde J_B\right \rangle=\beta_1 \eta_{AB}$, which is a degenerate invariant form for the Poincar\'e algebra (see equation \eqref{exoticITads}). The second term is the invariant tensor given in \eqref{invtensorGal}, while the third term is purely NR and does not follow from a relativistic invariant tensor upon contraction.}
\begin{equation}\label{invtensorGal}
\left\langle J  H\right \rangle  =-\kappa\,,
\end{equation}
which defines a degenerate invariant bilinear form for the Galilei algebra.

\subsection{The Poincar\'e $\ \oplus\ \mathfrak{u}(1)^2$ algebra}
In arbitrary dimensions, the contraction \eqref{redefinitionP} can be generalized by considering the direct sum of the Poincar\'e  algebra \eqref{Poincare} and a $\mathfrak{u}(1)$ generator $\tilde M$ \cite{Aldaya:1985plo} (see also \cite{Azcarraga:2011hqa}). This can be done by replacing the relation (\ref{Hredef}$\hskip -.1 truecm$) by
\begin{equation}\label{HMtilde}
\tilde H = \frac{1}{2}H +\varepsilon^2 M\,,\hskip 2truecm \tilde M =\frac{1}{2}H -\varepsilon^2 M\,.
\end{equation}
This contraction leads to the Bargmann algebra, which corresponds to the universal central extension of the Galilei algebra, and enlarges \eqref{GalileiD} by adding the commutation relation
\begin{equation}
[G_{a},P_{b}]=\delta_{ab} M\,.
\end{equation}
In three dimensions, one can go further and endow the Bargmann algebra with a second central extension by considering the direct sum of the Poincar\'e algebra \eqref{dualP} and two $\mathfrak{u}(1)$ generators $\tilde M$ and $\tilde S$ \cite{Bergshoeff:2016lwr} and replacing the relation (\ref{Jabredef}$\hskip -.1 truecm$) by
\begin{equation}\label{JStilde}
\tilde J = \frac{1}{2}J +\varepsilon^2 S\,,\hskip 2truecm \tilde S =\frac{1}{2}J -\varepsilon^2 S\,.
\end{equation}
Using \eqref{HMtilde}, \eqref{JStilde} and keeping the rescaling of the Galilean boosts and spatial translations generators\,\footnote{Like in the Galilei case, the scalings  do not coincide with ones used in \cite{Bergshoeff:2016lwr}. This fact is due an invariance under scaling of momenta of the Poincar\'e algebra, see footnote \ref{footnote1}.} as in (\ref{Gredef}$\hskip -.1 truecm$) and (\ref{Predef}$\hskip -.1 truecm$), the limit $\varepsilon\rightarrow\infty$ gives the
extended Bargmann algebra \cite{LL,BGGK}. Indeed, the corresponding inverse relations  are given by

\begin{multicols}{2}
\begin{subequations}\label{inverseredefinitionEB}
\setlength{\abovedisplayskip}{-13pt}
\allowdisplaybreaks
\begin{align}
 J &= \tilde J + \tilde S\,,\\[.2truecm]
 S &=\frac{1}{2\varepsilon^2}\big(\tilde J -\tilde S\big)\,,\\[.1truecm]
 G_a &= \frac{1}{\varepsilon}\tilde  G_a\,,\\[.1truecm]
 H &= \tilde H + \tilde M\,,\\[.2truecm]
 M &= \frac{1}{2\varepsilon^2}\big (\tilde H-\tilde M\big)\,,\\[.1truecm]
 P_a &= \frac{1}{\varepsilon} \tilde P_a\,.
\end{align}
\end{subequations}
\end{multicols}
\noindent
Using these relations, we find that, in the limit that $\varepsilon\rightarrow \infty$, the Poncar\'e $\oplus\ \mathfrak{u}(1)^2$ algebra reduces to the commutation relations of \eqref{Galilei} plus:
\begin{equation}
\label{extBargmann}
[G_a, P_b] =\epsilon_{ab}M\,,\hskip 2truecm
[G_a, G_b] = \epsilon_{ab}S\,,
\end{equation}
which corresponds to the double central extension of the Galilei algebra, where $M$ and $S$ are central charge generators.

The action for Poincar\'e $\oplus\ \mathfrak{u}(1)^2$ gravity is given by the same CS action \eqref{CSaction} as before, but with a gauge connection $A$ that now takes values in the Poincar\'e $\oplus\ \mathfrak{u}(1)^2$ algebra:
\begin{equation}\label{connectionPu12}
A = \Omega^A\tilde J_A + E^A\tilde P_A + a_1\tilde S + a_2 \tilde M \,.
\end{equation}
Accordingly, the invariant tensor \eqref{pairingR} has to be supplemented with the following invariant form for the Abelian generators:
\begin{equation}\label{pairingRU}
\left\langle \tilde M\tilde S\right\rangle = \alpha_2\,,
\end{equation}
 where $\alpha_2$ is an arbitrary parameter.
Using the relations
\eqref{inverseredefinitionEB}, the gauge connection $A$ can be expressed as
\begin{equation}
A = \omega J + \omega^a G_a +\tau H + e^a P_a + mM + sS\,,
\end{equation}
where the NR gauge fields are related to the relativistic ones as follows:
\begin{multicols}{2}
\begin{subequations}\label{gfcontractionPu12}
\setlength{\abovedisplayskip}{-13pt}
\allowdisplaybreaks
\begin{align}
\Omega^0&=\omega + \frac{1}{2\varepsilon^2} s\,,\\[.1truecm]
\Omega^a &= \frac{1}{\varepsilon}\omega^a\,,\\[.1truecm]
a_1 &= \omega - \frac{1}{2\varepsilon^2} s\,,\\[.1truecm]
E^0 &=\tau + \frac{1}{2\varepsilon^2} m\,,\\[.1truecm]
E^a &= \frac{1}{\varepsilon} e^a\,,\\[.1truecm]
a_2 &= \tau - \frac{1}{2\varepsilon^2} m\,.
\end{align}
\end{subequations}
\end{multicols}
\noindent
Using the above relations, the Poincar\'e $\oplus\ \mathfrak{u}(1)^2$ gravity  action \eqref{CSaction} takes the form
\begin{equation}\label{SPu1}
\begin{aligned}
S_{\textrm{Poincar\'e} \,\oplus\, \mathfrak{u}(1)^2}  = 2\int &\Bigg[(-\alpha_1+\alpha_2)\tau R(J) +\frac{\alpha_1}{\varepsilon^2} e^a R_a(G) \\
&-\frac{(\alpha_1+\alpha_2)}{2\varepsilon^2}\left(mR(J) + \tau R(S)\right)+\mathcal{O}(\varepsilon^{-4})\Bigg]\,,
\end{aligned}
\end{equation}
where, apart from \eqref{RJRG}, we have defined
\begin{equation}\label{RS}
R(S)=ds+\frac{1}{2}\epsilon_{ab}\,\omega^a \omega^b \,,
\end{equation}
which is the curvature 2-form associated to the central charge generator $S$
of the extended Bargmann algebra. We now consider two different set of values for the parameters $\alpha_1$ and $\alpha_2$.
\vskip .3truecm
\noindent {\bf Galilei gravity}
\vskip .1truecm

For general $\alpha_1 \ne \alpha_2$ we re-obtain, after taking the limit $\varepsilon\rightarrow \infty$, the Galilei gravity action \eqref{Gaction} constructed in the previous subsection.
\vskip .3truecm
\noindent {\bf Extended Bargmann gravity}
\vskip .1truecm

In the case that we consider
\begin{equation}\label{redefEB}
\alpha_1=\alpha_2=\varepsilon^2\kappa\,,
\end{equation}
we obtain an enhancement of Galilei gravity. In fact, the first term in \eqref{SPu1} vanishes and taking the limit $\varepsilon\rightarrow \infty$, we obtain the action for Extended Bargmann gravity \cite{Papageorgiou:2009zc,Bergshoeff:2016lwr,Hartong:2016yrf}
\begin{equation}\label{SEBG}
S_{\textrm{EBG}} =2\kappa \int \big[ e^a R_a(G) -mR(J) - \tau R(S)\big]\,.
\end{equation}
This result is consistent with
the known non-degenerate invariant bilinear form for the Extended Bargmann algebra:
\begin{multicols}{2}
\begin{subequations}\label{invtensorEBG}
\setlength{\abovedisplayskip}{-14pt}
\allowdisplaybreaks
\begin{align}
\left\langle J,M\right\rangle &=-\kappa\,,\\[.1truecm]
\left\langle S,H\right\rangle&=-\kappa \,,\\[.1truecm]
\left\langle G_a,P_b\right\rangle &= \kappa \delta_{ab}\,,
\end{align}
\end{subequations}
\end{multicols}\noindent
which follows from \eqref{inverseredefinitionEB}, \eqref{pairingRU} and \eqref{redefEB} in the limit $\varepsilon\rightarrow \infty$.
Note that the equations of motion of \eqref{SEBG} lead to the vanishing of all
curvatures.

\section{NR limits and the coadjoint Poincar\'e algebra}
\label{NRCP}

It is natural to ask oneself  if the NR
actions of \cite{Hansen:2018ofj,Hansen:2019vqf,Bergshoeff:2019ctr,Ozdemir:2019orp}
can be obtained from a relativistic action with an enlarged Poincar\'e symmetry algebra. If that is the case, what is the symmetry algebra?
In this section we address the question in the (2+1)-dimensional case and show how  the action of \cite{Ozdemir:2019orp}
can be obtained as the NR limit of a relativistic CS action, pretty much in the same way that the action for Extended Bargmann gravity is obtained as the limit of a relativistic CS action invariant under the Poincar\'e algebra. In order to do this, we extend  the algebras used in  the previous section as follows:
\begin{eqnarray}
\textrm{Poincar\'e}\hskip .5truecm &\rightarrow&\hskip .5truecm \textrm{coadjoint Poincar\'e}\,, \\[.2truecm]
\textrm{Poincar\'e} \oplus \mathfrak{u}(1)^2 \hskip .5truecm &\rightarrow&\hskip .5truecm \textrm{coadjoint Poincar\'e}  \oplus \mathfrak{u}(1)^2 \label{secondcase}\,.
\end{eqnarray}
As we will see, the relativistic symmetry behind this construction is the coadjoint Poincar\'e algebra, and it is only in the second case \eqref{secondcase} that we obtain the action of \cite{Ozdemir:2019orp}. We will now discuss these two cases separately.

\subsection{The Coadjoint Poincar\'e algebra}
In the following we will consider the following extension of the Poincar\'e algebra
\begin{multicols}{2}
\begin{subequations} \label{coadjointP}
\setlength{\abovedisplayskip}{-14pt}
\allowdisplaybreaks
\begin{align}
[\tilde J_{ A B},  \tilde P_{ C}] &=  2\eta_{ C[ B}\tilde P_{ A]}\,,\\[.1truecm]
[\tilde J_{ A B}, \tilde J_{ C D}] &= 4\eta_{[ A[ C}\tilde J_{ D] B]}\,,\\[.1truecm]
[\tilde J_{ A B},\tilde T_{ C}]  &= 2\eta_{ C[ B}\tilde T_{ A]}\, , \\[.1truecm]
[\tilde J_{ A B},\tilde S_{ C D}] &= 4\eta_{[ A[ C}\tilde S_{ D] B]}\, ,\\[.1truecm]
[\tilde S_{ A B},\tilde P_{ C}]  &= 2\eta_{ C[ B}\tilde T_{ A]}\, . \label{SPT}
\end{align}
\end{subequations}
\end{multicols}\noindent
We will refer to this algebra as coadjoint Poincar\'e algebra since it defines a consistent action of the Poincar\'e algebra on its dual space (see Appendix in \cite{Barducci:2019jhj}). However, as we will see in the following, the invariant tensor associated to this algebra is not the standard bilinear form that follows from the general double extension construction \cite{medina,Figueroa-OFarrill:1995opp}\footnote{We acknowledge Roberto Casalbuoni and Axel Kleinschmidt for discussions about general coadjoint actions.}. It is important to note that another definition of the coadjoint algebra
can be given in terms of the infinitesimal coadjoint representation associated to a semi-direct product group \cite{rawnsley1975,baguis1998,Barnich:2015uva},
where the existence of a non-degenerate invariant metric is guaranteed and corresponds to a particular case of a double extended algebra \cite{medina,Figueroa-OFarrill:1995opp}.
In the subsequent sections we will restrict to the (2+1)-dimensional case, where is straightforward to show that these two definitions mentioned above are isomorphic.


Now we divide the generators into space and time components by splitting the indices in the form $A=\{ 0,a \}$, which yields \eqref{splittingJP} together with
\begin{equation}\label{splittingST}
 \tilde S_{ A B}=\{ \tilde S_{0a} \equiv \tilde B_{a} \,, \; \tilde S_{ab}\}\, ,\hskip 2truecm
\tilde  T_{ A}=\{\tilde  T_0\equiv \tilde M \,,\; \tilde T_{a}\}\, ,
\end{equation}
%
%
%
%
%
%
%
%
and consider the following contraction of the coadjoint Poincar\'e algebra:
\begin{multicols}{2}
\begin{subequations} \label{contactioncoadP1}
\setlength{\abovedisplayskip}{-14pt}
\allowdisplaybreaks
\begin{align}
\tilde  J_{ab} &=  J_{ab}\,,\label{redefJcoad}\\[.2truecm]
 \tilde H &=  H\,,\label{redefHcoad}\\
\tilde G_{a} &=\frac{\varepsilon}{2}   G_{a}-\frac{\varepsilon^3}{2}   B_{a} \,, \label{redefGcoad}\\
\tilde P_{a} &= \frac{\varepsilon}{2}   P_{a}-\frac{\varepsilon^3}{2}   T_{a}\,, \label{redefPcoad}\\[.1truecm]
\tilde S_{ab} &= -\varepsilon^2 S_{ab}\,,\label{redefScoad}\\[.2truecm]
 \tilde M & =-\varepsilon^2 M\,,\label{redefMcoad}\\[.2truecm]
 \tilde B_{a} &=-\varepsilon   G_{a}-\varepsilon^3   B_{a} \,, \label{redefBcoad}\\[.2truecm]
\tilde T_{a} &= -\varepsilon   P_{a}- \varepsilon^3   T_{a} \label{redefTcoad}\,.
\end{align}
\end{subequations}
\end{multicols}\noindent
 %
 %
 %
Note that this contraction is different than the $k=1$ contraction of the coadjoint Poincar\'e algebra discussed in \cite{Barducci:2019jhj}. In the limit $\epsilon\rightarrow\infty$ this leads to the following commutation relations for the NR generators: 
\begin{multicols}{2}
\begin{subequations} \label{contractioncoadp}
\setlength{\abovedisplayskip}{-14pt}
\allowdisplaybreaks
\begin{align}
[   G_{a},   H]&=   P_{a}\, ,\\[.1truecm]
[   G_{a},   M]&=   T_{a}\, ,\\[.1truecm]
[   B_{a},   H]&=   T_{a}\, ,\\[.1truecm]
[   G_{a},   P_{b}]&=  \delta_{ab} M \, , \\[.1truecm]
[   G_{a},   G_{b}]&= S_{ab} \,,\\[.1truecm]
[   S_{ab},   G_{c}]&= 2\delta_{c[b}   B_{a]}\, ,\\[.1truecm]
[   S_{ab},   P_{c}]&= 2\delta_{c[b}   T_{a]}\, ,\\[.1truecm]
[   J_{ab},   S_{cd}]&=4\delta_{[a[c}   S_{d]b]}\, ,\\[.1truecm]
[   J_{ab},   J_{cd}]&=4\delta_{[a[c}   J_{d]b]}\, ,\\[.1truecm]
[   J_{ab},   X_{c}]&= 2\delta_{c[b}   X_{a]}\,,
\end{align}
\end{subequations}
\end{multicols}\noindent
where we have used the collective notation $ X_{a}= \left\{    G_{a},   P_{a},   B_{a},   T_{a} \right\} $. This corresponds to the algebra found by Hansen, Hartong and Obers in \cite{Hansen:2018ofj}, which can be obtained as a quotient of the infinite-dimensional NR expansion of the Poincar\'e algebra given in \cite{Hansen:2019vqf,Gomis:2019fdh}. Dividing out this algebra by the generators $\{B_a,T_a\}$ leads to a $D>3$ version of the Extended Bargmann algebra (see Appendix \ref{appendixA}).

Like in the Poincar\'e case, in three space-time dimensions we can dualize the generator of rotations as in \eqref{dualJ}, together with a similar definition for $\tilde S_{AB}$
\begin{equation}\label{dualS}
\tilde S_{AB}=\epsilon^{C}_{\;\;AB} \tilde S_{C}\,.
\end{equation}
This allows us to rewrite the coadjoint Poincar\'e algebra as \eqref{dualP} plus
\begin{equation} \label{coadjointP2+1}
[ \tilde J_A, \tilde S_B ]=\epsilon^{C}_{\;\;AB}\, \tilde S_C\,,\qquad
[ \tilde J_A, \tilde T_B ]=\epsilon^{C}_{\;\;AB}\, \tilde T_C\,,\qquad
[ \tilde S_A, \tilde P_B ]=\epsilon^{C}_{\;\;AB}\, \tilde T_C\,.
\end{equation}
In order to evaluate the NR limit, the equation \eqref{splitJA} has to be supplemented with
\begin{equation}
\tilde S_A=\left\{ \tilde S_0\equiv \tilde S\,, \; \tilde S_a\equiv \tilde B_a\right\}\,,
\end{equation}
while the definition \eqref{dualS} implies \eqref{dualNRJG} and
\begin{equation}\label{dualNRSB}
S_{ab}=-\epsilon_{ab}S\,,\hskip 2truecm B_a\rightarrow -\epsilon_a^{\;\;b}B_a\,.
\end{equation}
Using these redefinitions, the (2+1)-dimensional version of the algebra \eqref{contractioncoadp} takes the form of \eqref{Galilei} plus the following commutation relations:

\begin{multicols}{2}
\begin{subequations} \label{postNl2+1}
\setlength{\abovedisplayskip}{-14pt}
\allowdisplaybreaks
\begin{align}
[J,B_{a}]&=-\epsilon_{a}^{\;\;b}\, B_{b}\,\\[.1truecm]
[J,T_{a}]&=-\epsilon_{a}^{\;\;b}\, T_{b}\,,\\[.1truecm]
[S,G_{a}]&=-\epsilon_{a}^{\;\;b}\, B_{b}\,,\\[.1truecm]
[S,P_{a}]&=-\epsilon_{a}^{\;\;b}\, T_{b}\,,\\[.1truecm]
[G_{a},G_{b}]&=\epsilon_{ab}\,S\,,\\[.1truecm]
[M,G_{a}]&=-\epsilon_{a}^{\;\;b}\, T_{b}\,,\\[.1truecm]
[G_{a},P_{b}]&=\epsilon_{ab}\,M\,,\\[.1truecm]
[H,B_{a}]&=-\epsilon_{a}^{\;\;b}\, T_{b}\,.
\end{align}
\end{subequations}
\end{multicols}\noindent

We can construct a three-dimensional NR gravity theory with this symmetry by starting with the CS action \eqref{CSaction} invariant under the coadjoint Poincar\'e algebra, where now $A$ is a connection taking values on the coadjoint Poincar\'e algebra, i.e.
\begin{equation}\label{relconnection}
A = \Omega^A  \tilde J_A + E^A  \tilde P_A + \Sigma^A  \tilde S_A + L^A \tilde  T_A\,.
\end{equation}
We will consider the following invariant tensor:
\begin{equation}\label{invtensorcoadP}
\left\langle  \tilde J_A   \tilde T_B  \right\rangle= \gamma_1 \eta_{AB} \,,\quad
\left\langle  \tilde S_A  \tilde  P_B  \right\rangle= \gamma_1 \eta_{AB} \,,\quad
\left\langle  \tilde J_A   \tilde P_B  \right\rangle= \gamma_2 \eta_{AB} \,,
\end{equation}
where $\gamma_1$ and $\gamma_2$ are arbitrary parameters. This invariant tensor is non-degenerate for $\gamma_1\neq 0$.

The corresponding curvature two-form reads
\begin{equation}\label{curvaturecoadP}
R=dA+A^2=R^A(\tilde J) \,\tilde J_A+R^A(\tilde P)\,\tilde P_A+R^A(\tilde S) \,\tilde S_A+R^A(\tilde T) \,\tilde T_A\,,
\end{equation}
where $R^A(\tilde J)$ and $R^A(\tilde P)$ are given in \eqref{curvaturecompP} and
\begin{equation}\label{curvaturecompcoadP}
R^A(\tilde S)=d\Sigma^A+\epsilon^A_{\;\;BC}\Omega^B\Sigma^C \,,\hskip .5truecm
R^A(\tilde T)=d L^A+\epsilon^A_{\;\;BC}\left(\Omega^B L^C+\Sigma^B E^C\right) \,.
\end{equation}
The CS action for Coadjoint Poincar\'e gravity is given by
\begin{equation}
S_{\textrm{Coad-Poincar\'e}} =  2\gamma_1 \int \left[ E^A R_A(\tilde S)+L^A R_A(\tilde J)  \right]
+2\gamma_2 \int E^A R_A(\tilde J)\,.
\end{equation}
Note that the term with $\gamma_2$ gives the Einstein-Hilbert action in 2+1 dimensions.

We next study the contraction \eqref{contactioncoadP1} at the level of the CS action. Using the redefinitions \eqref{contactioncoadP1} the gauge connection \eqref{relconnection} can be written as

\begin{equation}
A= \omega J +\omega^a G_a + \tau H + e^a P_a + s S + b^a B_a + m M + t^a T_a \,,
\end{equation}
where the NR fields are related to the relativistic ones by
\begin{multicols}{2}
\begin{subequations} \label{contractionCoadP}
\setlength{\abovedisplayskip}{-13pt}
\allowdisplaybreaks
\begin{align}
\Omega^0&= \omega\,,\\[.2 truecm]
\Omega^a&= \frac{1}{\varepsilon} \omega^a -\frac{1}{\varepsilon^3} b^a \,,\\[.1 truecm]
\Sigma^0&= -\frac{1}{\varepsilon^2}s\,,\\[.1 truecm]
\Sigma^a&=- \frac{1}{2\varepsilon} \omega^a -\frac{1}{2\varepsilon^3} b^a \,,\\[.1 truecm]
E^0&= \tau \,,\\[.2 truecm]
E^a&= \frac{1}{\varepsilon}  e^a -\frac{1}{\varepsilon^3} t^a\,,\\[.1 truecm]
L^0&= -\frac{1}{\varepsilon^2}m \,,\\[.1 truecm]
L^a&= -\frac{1}{2\varepsilon}  e^a -\frac{1}{2\varepsilon^3} t^a\,,
\end{align}
\end{subequations}
\end{multicols}\noindent
and in terms of which the CS action takes the form:
\begin{equation}\label{actionCoadP}
\begin{aligned}
S_{\textrm{Coad-Poincar\'e}}=2 \int \Bigg[& -\gamma_2\tau R(J)+\frac{(-\gamma_1+\gamma_2)}{\varepsilon^2}e^a R_a (G)\\
&+\frac{(\gamma_1-\gamma_2)}{\varepsilon^2}\tau R(S)+\frac{\gamma_1}{\varepsilon^2}m R(J)+\frac{\gamma_2}{\varepsilon^2}\tau ds+ \mathcal{O}(\varepsilon^{-4})
\Bigg]\,.
\end{aligned}
\end{equation}
In the following, we consider 
two different set of values for the parameters $\gamma_1$ and $\gamma_2$ in this action.
\vskip .3truecm
\noindent {\bf Galilei gravity}
\vskip .1truecm

For general $\gamma_1 \ne \gamma_2$ we re-obtain, after taking the limit $\varepsilon\rightarrow \infty$, the Galilei gravity action \eqref{Gaction} constructed in the previous subsection.
\vskip .3truecm
\noindent {\bf Extended Bargmann gravity}
\vskip .1truecm

On the other hand, setting
\beq\label{thischoice}
\gamma_2=0 \,,\quad \gamma_1=-\kappa\varepsilon^2\,,
\eeq
we obtain the Extended Bargmann Gravity action of
\cite{Papageorgiou:2009zc,Bergshoeff:2016lwr,Hartong:2016yrf}:
\begin{equation}
 S_{\textrm{Extended-Bargmann}} = 2\kappa \int \Bigg[ e^a R_a (G)-\tau R(S)- m R(J)
\Bigg]\,.
\end{equation}
Substituting, for the choice of parameters \eqref{thischoice}, the expansion \eqref{contactioncoadP1} into the expression  \eqref{invtensorcoadP} for the invariant tensor leads the known invariant bilinear form of  Extended Bargmann Gravity given in equation \eqref{invtensorEBG}. Note that this invariant form is non-degenerate with respect to the Extended Bargmann algebra but it is degenerate with respect to the bigger algebra  \eqref{postNl2+1}.


We have not been able to define other sets of values of the
parameters leading to a finite action.
In particular, the three-dimensional counterpart of the NR gravity action proposed in \cite{Hansen:2018ofj}
can not be obtained as a NR limit of the CS action corresponding to the Coadjoint Poincar\'e algebra.


\subsection{The co-adjoint Poincar\'e\ $\oplus\ \mathfrak{u}(1)^2$ algebra}
\label{coadPplus2u1}

Similarly to what happens in the Galilean case, in arbitrary dimensions it is possible to generalize the contraction \eqref{contactioncoadP1} by considering the direct sum of the coadjoint Poincar\'e algebra \eqref{coadjointP} and a $\mathfrak{u}(1)$ generator $\tilde Y$. This is implemented by replacing the relations (\ref{redefHcoad}$\hskip -.15 truecm$) and (\ref{redefMcoad}$\hskip -.15 truecm$) by
\begin{equation}\label{HMYtilde}
\tilde H = \frac{1}{2}H -\varepsilon^4 Y\,,\qquad \tilde Y =\frac{1}{2}H +\varepsilon^4 Y\,,\qquad \tilde M =-\varepsilon^2 M-\varepsilon^4 Y\,.
\end{equation}
In fact, this leads to an extension of the algebra \eqref{contractioncoadp} by the commutation relations.
\begin{equation}
[G_{a},T_{b}]= \delta_{ab} Y\,,\hskip 2truecm
[B_{a},P_{b}]= \delta_{ab} Y\,.
\end{equation}

As it happens in the case of the Bargmann algebra, in three space-time dimensions one can generalize the previous result to include a second central extension by considering the direct sum of the coadjoint Poincar\'e algebra in 2+1 dimensions and two $\mathfrak{u}(1)$ generators $\tilde Y$ and $\tilde Z$. This is done by considering \eqref{HMYtilde} and replacing the relations (\ref{redefJcoad}$\hskip -.15 truecm$) and (\ref{redefScoad}$\hskip -.15 truecm$) by
\begin{equation}\label{JSZtilde}
\tilde J = \frac{1}{2}J -\varepsilon^4 Z\,,\qquad \tilde Z =\frac{1}{2}J +\varepsilon^4 Z\,,
\qquad \tilde S =-\varepsilon^2 S-\varepsilon^4 Z\,.
\end{equation}
Using \eqref{HMYtilde}, \eqref{JSZtilde} and keeping the definitions (\ref{redefGcoad}$\hskip -.1 truecm$)--(\ref{redefTcoad}$\hskip -.1 truecm$) for the generators $\tilde G_a$, $\tilde P_a$, $\tilde B_a$ and $\tilde T_a$, the limit $\varepsilon\rightarrow\infty$ yields the bosonic algebra presented in \cite{Ozdemir:2019orp} in the context of a novel supersymmetric NR gravity theory. This can be seen by considering the inverse relations

\begin{multicols}{2}
\begin{subequations} \label{inversenr}
\setlength{\abovedisplayskip}{-13pt}
\allowdisplaybreaks
\begin{align}
   J&= \tilde J+\tilde Z\,,\\[.2truecm]
  G_{a}&=\dfrac{1}{\varepsilon} \left( \tilde G_{a}-\dfrac{1}{2}\tilde B_{a} \right)\,,\\[.1truecm]
   S&= -\dfrac{1}{\varepsilon^2}  \left( \tilde S -\frac{1}{2} \tilde J+\frac{1}{2} \tilde Z\right)\,,\\[.1truecm]
  B_{a}&=-\dfrac{1}{\varepsilon^3} \left( \tilde G_{a}+\dfrac{1}{2}\tilde B_{a} \right)\,, \\[.1truecm]
  Z&= -\dfrac{1}{2\varepsilon^4}\left( \tilde J -\tilde Z \right)\,, \\[.1truecm]
  H&= \tilde H +\tilde Y\,, \\[.2truecm]
  P_{a} &= \dfrac{1}{\varepsilon} \left( \tilde P_{a}-\dfrac{1}{2}\tilde T_{a} \right) \,,\\[.1truecm]
  M &=-\dfrac{1}{\varepsilon^2}  \left( \tilde M -\frac{1}{2} \tilde H +\frac{1}{2} \tilde Y \right)\,,\\[.1truecm]
  T_{a} &=- \dfrac{1}{\varepsilon^3} \left( \tilde P_{a}+\dfrac{1}{2}\tilde T_{a} \right) \,,\\[.1truecm]
  Y &=-\dfrac{1}{2\varepsilon^4} \left( \tilde H -\tilde Y \right)\,.
\end{align}
\end{subequations}
\end{multicols}\noindent
One can see that in the limit $\varepsilon\rightarrow\infty$, these generators satisfy the commutation relations \eqref{Galilei} and \eqref{postNl2+1} plus
\begin{equation}\label{extYZ3D}
[G_{a},B_{b}]=\epsilon_{ab}\, Z\,,\qquad
[G_{a},T_{b}]=\epsilon_{ab}\,Y\,,\qquad
[B_{a},P_{b}]=\epsilon_{ab}\,Y\,,
\end{equation}
which define the double central extension
\cite{medina,Figueroa-OFarrill:1995opp,Matulich:2019}
of the three-dimensional version of the algebra \eqref{contractioncoadp}.
Moreover, in the free algebra construction \cite{Gomis:2019fdh}, this algebra can be obtained as the three-dimensional version of the a suitable quotient of an infinite-dimensional expansion of the Poincar\'e algebra (see Appendix \ref{appendixA})

The corresponding NR gravity theory can be obtained as a contraction of a CS action invariant under the coadjoint Poincar\'e $\oplus\ \mathfrak{u}(1)^2$ algebra. This theory can be defined by supplementing the connection \eqref{relconnection}  with  two extra Abelian gauge fields which we denote by $a_1$ and $a_2$, i.e.
\begin{equation}\label{relconnection2}
A = \Omega^A  \tilde J_A + E^A  \tilde P_A + \Sigma^A  \tilde S_A + L^A \tilde  T_A + a_1 \tilde Z+ a_2 \tilde Y \,.
\end{equation}
Using \eqref{inversenr} the connection can be expressed as
\beq
A= \omega J +\omega^a G_a + \tau H + e^a P_a + s S + b^a B_a + m M + t^a T_a +y Y + z  Z\,,
\end{equation}
where the NR fields are related to the relativistic ones by

\begin{multicols}{2}
\begin{subequations}\label{relgaugefields2}
\setlength{\abovedisplayskip}{-12pt}
\allowdisplaybreaks
\begin{align}
\Omega^0&= \omega+\frac{1}{2\varepsilon^2} s-\frac{1}{2\varepsilon^4} z\,,\\[.1truecm]
\Omega^a&= \frac{1}{\varepsilon} \omega^a -\frac{1}{\varepsilon^3} b^a \,,\\[.1truecm]
\Sigma^0&= -\frac{1}{\varepsilon^2}s\,,\\[.1truecm]
\Sigma^a&=- \frac{1}{2\varepsilon} \omega^a -\frac{1}{2\varepsilon^3} b^a \,,\\[.1truecm]
a_1&= \omega-\frac{1}{2\varepsilon^2} s+\frac{1}{2\varepsilon^4} z\,,\\[.1truecm]
E^0&= \tau+\frac{1}{2\varepsilon^2} m-\frac{1}{2\varepsilon^4} y \,,\\[.1truecm]
E^a&= \frac{1}{\varepsilon}  e^a -\frac{1}{\varepsilon^3} t^a\,,\\[.1truecm]
L^0&= -\frac{1}{\varepsilon^2}m \,,\\[.1truecm]
L^a&= -\frac{1}{2\varepsilon}  e^a -\frac{1}{2\varepsilon^3} t^a\,,\\[.1truecm]
a_2&= \tau-\frac{1}{2\varepsilon^2} m+\frac{1}{2\varepsilon^4} y \,.
\end{align}
\end{subequations}
\end{multicols}\noindent
In order to explicitly compute the CS gravity action, the invariant tensor \eqref{invtensorcoadP} has to be supplemented with
\begin{equation}\label{invtensorcoadPu12}
\left\langle \tilde Y\, \tilde Z\right\rangle= \gamma_3\,,
\end{equation}
where we have introduced an extra arbitrary parameter $ \gamma_3$.
The corresponding CS action then takes the form
\begin{equation}\label{ScoadPu1}
\begin{aligned}
S_{\textrm{coad-Poicar\'e}\,\oplus\,\mathfrak{u}(1)^2}=  2 \int& \Bigg[(-\gamma_2+\gamma_3)\tau R(J) +\frac{(2\gamma_1-\gamma_2-\gamma_3)}{2\varepsilon^2}m R(J)+\frac{(\gamma_1-\gamma_2)}{\varepsilon^2}\tau R(S)
\\
& + \frac{(\gamma_2-\gamma_3)}{2\varepsilon^2}\tau ds+\frac{(-\gamma_1+\gamma_2)}{\varepsilon^2}e^a R_a (G) +\frac{(\gamma_2+\gamma_3)}{2\varepsilon^4} y R(J) +\frac{\gamma_2}{\varepsilon^4}\tau R(Z)
\\
& +\frac{(3\gamma_1-\gamma_2)}{2\varepsilon^4}m R(S) +\frac{3(\gamma_1-\gamma_2)}{2\varepsilon^4}\epsilon_{ab} e^a\omega^b s
 +\frac{(-\gamma_2+\gamma_3)}{2\varepsilon^4}\tau dz
\\
&+\frac{(-2\gamma_1+\gamma_2+\gamma_3)}{4\varepsilon^4}m ds
-\frac{\gamma_2}{\varepsilon^4} \left( e^a R_a (B) + t^a R_a (G) \right)
+ \mathcal{O}(\varepsilon^{-6})
\Bigg]
 \,,
\end{aligned}
\end{equation}
where we have used \eqref{RJRG}, \eqref{RS} and defined
\begin{equation}\label{RZRB}
R(Z)=dz+\epsilon_{ab}\,\omega^a b^b
 \,,\hskip 1.5 truecm
R^a(B)=db^a +\epsilon^a_{\;\;b}\left(s\,\omega^b + \omega\, b^b \right) \,.
\end{equation}
We now consider 
three different set of values for the parameters.
\vskip .3truecm

\noindent \textbf{Galilei  gravity}
\vskip .1truecm

For general $\gamma_2 \ne \gamma_3$ we re-obtain, after taking the limit $\varepsilon\rightarrow \infty$, the Galilei gravity action \eqref{Gaction}.
\vskip .3truecm

\noindent \textbf{Extended Bargmann gravity}
\vskip .1truecm


If we set $\gamma_2=\gamma_3=0$ and implement the rescaling $\gamma_1=-\varepsilon^2\kappa$, the action \eqref{ScoadPu1} leads to no divergent terms, allowing us to recover, in the limit that $\varepsilon\rightarrow\infty$, the Extended Bargmann gravity action of \cite{Papageorgiou:2009zc,Bergshoeff:2016lwr,Hartong:2016yrf} given in \eqref{SEBG}.
The same result can be recovered by setting $\gamma_1=0$ and $\gamma_2=\gamma_3=\varepsilon^2\kappa$. This second option was to be expected since the invariant tensor \eqref{invtensorcoadP} matches the one of the Poincar\'e $\oplus\ \mathfrak{u}(1)^2$ algebra \eqref{invtensorcoadPu12} when $\gamma_1=0$. Similarly, the corresponding CS action \eqref{ScoadPu1} then reduces  to the Einstein Hilbert action in three-dimensions plus two $\mathfrak{u}(1)$ fields. Note that this  choice yields a degenerate  invariant tensor for the Coadjoint-Poincar\'e algebra.
\vskip .3truecm

\noindent \textbf{Enhanced  Bargmann Gravity}
\vskip .1truecm

Setting
\begin{equation}\label{condalphas}
\gamma_1=\gamma_2=\gamma_3=-\varepsilon^4\kappa\,,
\end{equation}
and taking the limit $\varepsilon\rightarrow\infty$ we obtain the following enhancement of Extended Bargmann gravity
\begin{equation}
S_{\textrm{Enhanced-Bargmann}}=2\kappa \int \left(
e^a R_a (B)+t^a R_a (G)-\tau R(Z)- y R(J)- m R(S)   \right)
 \,,
\end{equation}
which is precisely the action that has been studied  in \cite{Ozdemir:2019orp}.

Substituting \eqref{inversenr} and \eqref{condalphas} into the invariant tensor \eqref{invtensorcoadPu12} and taking the limit that $\varepsilon\rightarrow\infty$ reduces to the following NR invariant tensor
\begin{multicols}{2}
\begin{subequations}
\setlength{\abovedisplayskip}{-15pt}
\allowdisplaybreaks
\begin{align}
\left\langle S,M\right\rangle&= \kappa\,,\\[.1truecm]
\left\langle Z,H\right\rangle&= \kappa\,,\\[.1truecm]
\left\langle J,Y\right\rangle&=\kappa \,,\\[.1truecm]
\left\langle G_a,T_b\right\rangle &=-\kappa\delta_{ab}\,,\\[.1truecm]
\left\langle B_a,P_b\right\rangle &=-\kappa\delta_{ab}\,.
\end{align}
\end{subequations}
\end{multicols}\noindent
which is non-degenerate.


\section{NR limits and the Coadjoint AdS algebra}
\label{NRCAdS}

The coadjoint AdS algebra in $D$ dimensions can be obtained by supplementing the coadjoint Poincar\'e commutation relations \eqref{coadjointP} with
\begin{equation}\label{coad1}
[\tilde P_{ A},\tilde P_{ B}] = \frac{1}{\ell^2} \tilde J_{ A B}\,, \hskip 2truecm [\tilde P_{ A},\tilde T_{ B}] = \frac{1}{\ell^2} \tilde S_{ A B}\, .
\end{equation}
where $\ell$ is the AdS radius. Naturally, in 2+1 dimensions, this algebra can be written as commutation relations \eqref{dualP}, \eqref{coadjointP2+1} and \eqref{coad1} with $\tilde J_{AB}=\epsilon_{AB}\tilde J$,  $\tilde S_{AB}=\epsilon_{AB}\tilde S$. By adding two $\mathfrak u(1)$ generators $\tilde Y$ and $ \tilde Z$, one can consider the contraction defined by \eqref{inversenr}. This defines a NR limit of coadjoint AdS$_3 \oplus \mathfrak u (1)^2$ given by the commutation relations of \eqref{Galilei} and \eqref{postNl2+1} plus
\begin{multicols}{2}
\begin{subequations} \label{NRL2}
\setlength{\abovedisplayskip}{-13pt}
\allowdisplaybreaks
\begin{align}
[H,P_{a}]&=-\frac{1}{\ell^2}\epsilon_{a}^{\;\;b}\, G_{b}\,,\\[.1truecm]
[M,P_{a}]&=-\frac{1}{\ell^2}\epsilon_{a}^{\;\;b}\, B_{b}\,,\\[.1truecm]
[H,T_{a}]&=-\frac{1}{\ell^2}\epsilon_{a}^{\;\;b}\, B_{b}\,,\\[.1truecm]
[P_{a},P_{b}]&=\frac{1}{\ell^2}\epsilon_{ab}\,S\,,\\[.1truecm]
[P_{a},T_{b}]&=\frac{1}{\ell^2}\epsilon_{ab}\,Z\,,
\end{align}
\end{subequations}
\end{multicols}\noindent
which agrees with the result of~\cite{Gomis:2019nih}.

The corresponding CS action \eqref{CSaction} follows from considering a connection $A$ of the form \eqref{relconnection} (now taking values in the coadjoint AdS$_3$ algebra) and the invariant tensor \eqref{invtensorcoadP}. In this case, the components $R^A[\tilde J] $and $R^A[\tilde S]$ of the curvature two-form \eqref{curvaturecoadP} have an extra term proportional to the cosmological constant:
\beq
\begin{aligned}
&R^A[\tilde J]=d\Omega^A+\frac{1}{2}\epsilon^A_{\;\;BC}\left(\Omega^B\Omega^C +\frac{1}{\ell^2} E^B E^C\right) \,,\\[6pt]
&R^A[\tilde S]=d\Sigma^A+\epsilon^A_{\;\;BC}\left(\Omega^B\Sigma^C +\frac{1}{\ell^2} E^B L^C\right)\,,
\end{aligned}
\eeq
while $R^A[\tilde P] $ and $R^A[\tilde T]$ have the same form given in \eqref{curvaturecompP} and \eqref{curvaturecompcoadP}, respectively. Evaluating the CS action leads to
\begin{equation}\label{CSactioncoadAdS}
\begin{aligned}
S_{CS}&=  2\gamma_2 \int \left[ E^A R_A[\tilde J] -\frac{1}{3\ell^2} \epsilon_{ABC}E^A E^B E^C \right]  \\
&+ 2\gamma_1 \int \left[ E^A R_A[\tilde S]+L^A R_A[\tilde J] -\frac{1}{\ell^2} \epsilon_{ABC} E^A E^B L^C \right]\,.
\end{aligned}
\end{equation}

Remarkably, the coadjoint AdS$_3$ algebra can be written as the direct sum of two $\mathfrak{iso}(2,1)$ algebras.
\beq\label{twopoincare}
\left[ \mathcal J^\pm_A, \mathcal J^\pm_A\right]=\epsilon^{C}_{\;\;AB}\, \mathcal J^\pm_C\,,\quad\left[ \mathcal J^\pm_A, \mathcal P^\pm_B\right]=\epsilon^{C}_{\;\;AB}\, \mathcal P^\pm_C\,,\quad\left[ \mathcal P^\pm_A, \mathcal P^\pm_B\right]=0\,,
\eeq
where these two sets Poincar\'e generators are related to the original basis $\{  \tilde J_A ,  \tilde P_A,  \tilde S_A,  \tilde T_A\}$ as
\beq\label{chiralbasis2xpoincare}
 \mathcal J^\pm_A=\frac{1}{2}\left(\tilde J_A\pm\ell \tilde P_A\right)\,,\hskip 2.5 truecm  \mathcal P^\pm_A=\frac{1}{2}\left(\frac{1}{\ell} \tilde S_A\pm \tilde T_A\right)\,.
\eeq
This result is a generalization of the well known result that for the pure AdS$_3$ case,
$\mathfrak{so}(2,2)=\mathfrak{so}(2,1)\oplus \mathfrak{so}(2,1)$. This can also be understood from the fact that the Poincar\'e algebra in 2+1 dimensions $\mathfrak{iso}(2,1)$ is isomorphic to the coadjoint $\mathfrak{so}(2,1)$ algebra \cite{Barnich:2015tba}.

The action \eqref{CSactioncoadAdS} can be alternatively expressed in terms of two independent sets of gauge fields that follow from the redefinition of the Lie algebra generators \eqref{chiralbasis2xpoincare}, i.e.
\beq
E_\pm ^A =  \ell\Sigma^A \pm  L^A \,,\hskip 2.5 truecm
\Omega_\pm ^A=  \Omega^A \pm \frac{1}{\ell} E^A
\eeq
In this case the connection can be written in terms of the Poincar\'e generators \eqref{twopoincare} as
\beq
A = A^+ + A^-\,,\hskip 2.5 truecm A^\pm=\Omega_\pm^A \mathcal J^\pm_A + E_\pm^A \mathcal P^\pm_A\,.
\eeq
Similarly, the curvature takes the simple form
\beq
R=R^+ + R^- \,,\hskip 1.5 truecm R^\pm=dA^\pm+A^{\pm\,2}=R_\pm^A[\mathcal J] \,\mathcal J^\pm_A+R_\pm^A[\mathcal P] \,\mathcal P^\pm_A \,,
\eeq
where
\beq
R_\pm^A[\mathcal J^\pm]=d\Omega_\pm^A+\frac{1}{2}\epsilon^A_{\;\;BC}\Omega_\pm^B\Omega_\pm^C  \,,\hskip 1.5 truecm
R_\pm^A[\mathcal P^\pm]=d E_\pm^A+\epsilon^A_{\;\;BC} \Omega_\pm^B E_\pm^C\,.\\[6pt]
\eeq
In this chiral basis the invariant tensor \eqref{invtensorcoadP} reads
\beq\label{invtcoadjointads}
\left\langle   \mathcal J^\pm_A   \mathcal P^\pm_B  \right\rangle=\pm \frac{ \gamma_1}{2}\eta_{AB} \,,\hskip 2.5 truecm
\left\langle  \mathcal J^\pm_A    \mathcal J^\pm_B  \right\rangle=\pm \frac{\ell \gamma_2}{2}\eta_{AB} \,,
\eeq
which means the CS action splits as
\begin{equation}\label{chiraldecaction}
S_{CS}[A]=S[A^+]-S[A^-]\,,
\end{equation}
where
\beq\label{CSchiralR}
S[A^\pm]= \gamma_1 \int E^A_\pm R_A^\pm[\mathcal J^\pm]\,+\frac{\ell \gamma_2}{2} \int \left[ \eta_{AB}\Omega ^A_\pm d \Omega^B_\pm+\frac{1}{3}\epsilon_{ABC} \Omega ^A_\pm \Omega ^B_\pm \Omega ^C_\pm  \right]\,.
\eeq
We can consider $\gamma_2=0$ and interpret the coadjoint AdS$_3$ gravity action as the sum of two Einstein-Hilbert terms without interactions. Making this choice it is clear that one could add one $\mathfrak u(1)$ generator to each copy of Einstein gravity and define the NR limit of coadjoint AdS$_3 \oplus \mathfrak u (1)^2$ gravity in three dimensions as two copies of the Bargmann algebra by following the approach of \cite{Papageorgiou:2009zc,Bergshoeff:2016lwr} on each independent chiral sector. Nevertheless, this would lead to a degenerate invariant form after performing the NR limit\footnote{Furthermore, it is interesting to note that one could use the action \eqref{CSchiralR} with $\gamma_2=0$ and add two Abelian fields to each copy of the Einstein-Hilbert action, which would lead to two copies of Extended Bargmann gravity as the NR limit of coadjoint AdS$_3\oplus \mathfrak u(1)^4$ gravity and to a non-degenerate NR invariant tensor.}.

However, as we will see in the following, the CS term for $\Omega ^A$ in \eqref{CSchiralR} will be important to define a different NR limit of coadjoint AdS$_3$ gravity that connects with the results previously shown in the coadjoint Poincar\'e case.

\subsection{NR limit of Coadjoint AdS$_3\oplus \mathfrak u (1)^2$ gravity}

The NR limit of coadjoint AdS$_3$ gravity in three dimensions can be studied in the same way as we previously did in the case of vanishing cosmological constant. Before starting the analysis it is important to recall what happens in the case of AdS$_3$ invariant CS gravity, where the Einstein Hilbert action \eqref{Poincaregrav} is modified in the form
\begin{equation}\label{AdSgrav}
S_{\textrm{AdS}}=2\alpha_1 \int \left( E^A R^B (\tilde J) -\frac{1}{\ell^2} \epsilon_{ABC} E^A E^B E^C\right)\,.
\end{equation}
Now the curvature component $R^A (\tilde J)$ is not given by \eqref{curvaturecompP}, but by
\begin{equation}
R^A(\tilde J)=d\Omega ^A+\frac{1}{2}\epsilon^{A}_{\;\; BC} \left( \Omega^B \Omega^C+\frac{1}{\ell^2}  E^B E^C\right) \,,
\end{equation}
while the torsion 2-form $R^A(\tilde P)$ keeps the same form as in \eqref{curvaturecompP}.

The contraction of the AdS algebra gives the Newton-Hooke algebra \cite{Bacry:1968zf}. However, the first term of the expansion of CS gravity corresponds to  Galilei gravity \eqref{Gaction} exactly as in the Poincar\'e case. The reason is that, in 2+1 dimensions, $R(J)=d\omega$ also in the Newton-Hooke case. The next step is to add two $\mathfrak u(1)$ generators, which requires to use the connection \eqref{connectionPu12}  and the invariant tensor \eqref{pairingR} together with \eqref{pairingRU}. At the level of the action, this implies that one has to add a term of the form
\begin{equation}\label{u12action}
 \int \left( a_1 d a_2+ a_2 d a_1 \right)\,,
\end{equation}
with global factor $\alpha_2$ to the action \eqref{AdSgrav}. Then, for $\varepsilon\rightarrow\infty$, the contraction \eqref{gfcontractionPu12} leads to Galilean gravity when $\alpha_1\neq\alpha_2$, and to Extended Bargmann--Newton--Hooke gravity \cite{Papageorgiou:2010ud,Hartong:2016yrf}
\begin{equation}\label{extNHaction}
 S_{\textrm{Extended-BNH}} = 2\kappa \int \Bigg[ e^a R_a (G)-\tau R(S)- m R(J)+\frac{1}{\ell^2}\epsilon_{ab}\tau e^a e^b
\Bigg]
\end{equation}
for $\alpha_1=\alpha_2=\varepsilon^2\kappa$. The corresponding invariant tensor in this case is non-degenerate and given by \eqref{invtensorEBG}. A more general non-degenerate invariant tensor for the Extended Bargmann--Newton--Hooke algebra, which becomes degenerate in the Extended Bargmann limit, comes from considering the relativistic invariant tensor for AdS
\begin{equation}\label{exoticITads}
\left\langle \tilde J_A \tilde J_B\right\rangle=\beta_1 \eta_{AB}\,,\hskip 2.5 truecm \left\langle \tilde P_A \tilde P_B\right\rangle=\frac{\beta_1}{\ell^2}\eta_{AB}
\end{equation}
in the contraction \eqref{inversenr}. We have not considered this case here.

Now that we have reviewed the well-known AdS case, we turn our attention to the novel case  of the coadjoint AdS$_3$ algebra. As it happens in the coadjoint Poincar\'e case, expressing the action \eqref{CSactioncoadAdS} in terms of the NR fields \eqref{contractionCoadP} does not lead to new NR actions. Indeed, Galilean gravity \eqref{Gaction} is recovered for $\gamma_2\neq\gamma_3$ and Extended Bargmann--Newton--Hooke gravity \eqref{extNHaction} follows from choosing $\gamma_1=-\varepsilon^2\kappa\,,\;\gamma_2=0$. Also, these CS theories are associated to degenerate invariant tensors for the coadjoint AdS$_3$ algebra.

As done in Section \ref{coadPplus2u1}, in order to obtain a new NR gravity action from the coadjoint AdS$_3$ algebra, we incorporate two Abelian fields into the theory by considering a connection of the form \eqref{relconnection2} and the invariant tensor formed by\eqref{invtensorcoadP} and \eqref{invtensorcoadPu12}. This leads to the action \eqref{CSactioncoadAdS} supplemented with the term \eqref{u12action} with global factor $\gamma_3$. Expressing the action in terms of NR gauge fields using \eqref{relgaugefields2}, one can consider the same three different sets of values for the parameters $\gamma_1$, $\gamma_2$ and $\gamma_3$ considered in Section \ref{coadPplus2u1}:
\vskip .3truecm

\noindent \textbf{Galilei  gravity}
\vskip .1truecm

In the case $\gamma_2 \ne \gamma_3$, the limit $\varepsilon\rightarrow \infty$ leads to the action for Galilei gravity \eqref{Gaction}.
\vskip .3truecm

\noindent \textbf{Extended Bargmann--Newton--Hooke gravity}
\vskip .1truecm


The choice $\gamma_2=\gamma_3=0$ together with the rescaling $\gamma_1=-\varepsilon^2\kappa$, leads the action for Extended Bargmann--Newton--Hooke gravity \eqref{extNHaction}. This result can also be obtained by setting $\gamma_1=0$ and $\gamma_2=\gamma_3=\varepsilon^2\kappa$.

\vskip .3truecm

\noindent \textbf{Enhanced  Bargmann--Newton--Hooke Gravity}
\vskip .1truecm

Choosing $\gamma_1=\gamma_2=\gamma_3=-\varepsilon^4\kappa$ and taking the limit $\varepsilon\rightarrow\infty$ yields the action
\begin{equation}\label{EBNHgrav}
\begin{aligned}
S_{\textrm{Enhanced-BNH}}=&2\kappa \int \Bigg(
e^a R_a (B)+t^a R_a (G)-\tau R(Z)\\
&\hskip 1.5 truecm - y R(J)- m R(S) +\frac{1}{\ell^2}\epsilon_{ab}\left(me^a e^b +2 \tau e^a t^b\right)  \Bigg)
 \,,
 \end{aligned}
\end{equation}
which will be referred to as Enhanced Bargmann--Newton--Hooke gravity, and has been previously constructed in \cite{Gomis:2019nih,Concha:2019dqs}.

\subsection{NR limit of Coadjoint AdS$_3\oplus \mathfrak u (1)^2$ gravity in chiral basis}

The NR contraction \eqref{contactioncoadP1} applied to the coadjoint AdS$_3$ algebra can be alternatively worked out in the chiral basis by relabelling the relativistic generators \eqref{chiralbasis2xpoincare} in the form
\begin{equation}\label{relabelchiral}
\mathcal J^\pm_A =\left\{ \mathcal J^\pm_0\equiv \mathcal J^\pm,\mathcal J^\pm_a\equiv \mathcal G^\pm_a\right\} \,,\hskip 2.0 truecm
\ell\mathcal P^\pm_A=\left\{ \ell\mathcal P^\pm_0\equiv \mathcal S^\pm, \ell\mathcal P^\pm_a\equiv \mathcal B^\pm_a\right\}\,. \\
\end{equation}

\noindent By adding two $\mathfrak{u}(1)$ generators $\mathcal Z^\pm$ to the coadjoint AdS${}_3$ algebra, we can define the following contraction for two copies of $\mathfrak{iso}(2,1)\oplus \mathfrak{u}(1)$:

\begin{multicols}{2}
\begin{subequations} \label{chiralcontraction}
\setlength{\abovedisplayskip}{-14pt}
\allowdisplaybreaks
\begin{align}
\mathcal J^\pm&=\dfrac{1}{2}   J^\pm-\varepsilon^4   Z^\pm \,, \\[.1truecm]
\mathcal G^\pm_{a}&= \frac{\varepsilon}{2} \,G^\pm_{a} -\frac{\varepsilon^3}{2} \,B^\pm_{a}\,, \\[.2truecm]
\mathcal S^\pm &= -\varepsilon^2 S^\pm-\varepsilon^4   Z^\pm\,,\\[.1truecm]
\mathcal Z^\pm &=\dfrac{1}{2}   J^\pm+\varepsilon^4   Z^\pm\,,\\[.2truecm]
\mathcal B^\pm_{a} &=-\varepsilon \,G^\pm_{a} -\varepsilon^3\,B^\pm_{a}\,.\\ \nonumber
\end{align}
\end{subequations}
\end{multicols}\noindent
Inverting this change of basis leads to the following expression for the NR generators
\begin{multicols}{2}
\begin{subequations}
\setlength{\abovedisplayskip}{-12pt}
\allowdisplaybreaks
\begin{align}
J^\pm&= \mathcal  J^\pm+\mathcal  Z^\pm \,,\\[.2truecm]
G^\pm_{a}&=\dfrac{1}{\varepsilon} \left( \mathcal  G^\pm_{a} -\frac{1}{2} \mathcal  B^\pm_{a} \right)\,,\\
S^\pm&=-\dfrac{1}{\varepsilon^2} \left( \mathcal  S^\pm-\frac{1}{2}\mathcal J^\pm +\frac{1}{2}\mathcal  Z^\pm \right)\,,\\
Z^\pm &=-\dfrac{1}{2\varepsilon^4} \left(\mathcal  J^\pm-\mathcal  Z^\pm \right)\,, \\
B^\pm_{a}&=- \dfrac{1}{\varepsilon^3}\left( \mathcal G^\pm_{a}+\frac{1}{2} \mathcal  B^\pm_{a}  \right) \,,\\ \nonumber
\end{align}
\end{subequations}
\end{multicols}\noindent
which satisfy the commutation relations
\begin{multicols}{2}
\begin{subequations} \label{2xchiralalgebra}
\setlength{\abovedisplayskip}{-14pt}
\allowdisplaybreaks
\begin{align}
[ J^\pm,  G^\pm_{a}]&=-\epsilon_{a}^{\;\;b}\,   G^\pm_{b}\,,\\[.1truecm]
[  J^\pm,  B^\pm_{a}]&=-\epsilon_{a}^{\;\;b}\,   B^\pm_{b}\,,\\[.1truecm]
[  S^\pm,  G^\pm_{a}]&=-\epsilon_{a}^{\;\;b}\,   B^\pm_{b}\,,\\[.1truecm]
[  G^\pm_{a},  B^\pm_{b}]&=\epsilon_{ab}\,  Z^\pm\,,\\[.1truecm]
[  G^\pm_{a},  G^\pm_{b}]&=\epsilon_{ab}\,  S^\pm\,.\\ \nonumber
\end{align}
\end{subequations}
\end{multicols}\noindent
The Enhanced Bargmann--Newton--Hooke symmetry given by the commutation relations \eqref{Galilei}, \eqref{postNl2+1} and \eqref{NRL2} can be recovered from \eqref{2xchiralalgebra} by defining
\begin{multicols}{2}
\begin{subequations} \label{nonCintermsofC}
\setlength{\abovedisplayskip}{-10pt}
\allowdisplaybreaks
\begin{align}
J&=J^+ +J^- \,,\\[.3truecm]
G_a &= G^+_a + G^-_a\,,\\[.3truecm]
S&=S^+ +S^- \,,\\[.3truecm]
B_a &= B^+_a + B^-_a\,,\\[.3truecm]
 Z&= Z^+ +Z^- \,,\\[.1truecm]
H&=\frac{1}{\ell}\left( J^+ - J^- \right)\,,\\
P_a& =\frac{1}{\ell}\left( G^+_a - G^-_a\right)\,,\\
M&=\frac{1}{\ell}\left( S^+ - S^- \right)\,,\\
T_a& =\frac{1}{\ell}\left( B^+_a - B^-_a\right)\,,\\
 Y&= \frac{1}{\ell}\left( Z^+ - Z^- \right)\,.
\end{align}
\end{subequations}
\end{multicols}\noindent

The algebra \eqref{2xchiralalgebra} can be alternatively obtained as a finite NR expansion of the $\mathfrak{so}(2,1)$ algebra
\begin{equation}
\left[ \tilde J_A, \tilde J_B\right]=\epsilon^C_{\;\;AB}\tilde J_C
\end{equation}
by means of the semigroup $S_E^{(N)}$ for $N=4$ \cite{Izaurieta:2006zz}. This, in turn, corresponds to a suitable quotient of an infinite-dimensional expansion of the Lorentz algebra in three dimensions \cite{Gomis:2019nih}. Moreover, since  $\mathfrak{so}(2,1)$ also defines AdS$_2$, we can interpret \eqref{2xchiralalgebra} as an expansion of the AdS algebra in 1+1 dimensions. Thus, this result generalizes the fact that the Extended Bargmann--Newton--Hooke algebra in three dimensions can be written as two copies of the Bargmann--Newton--Hooke algebra in two dimensions \cite{Alvarez:2007ys,Alvarez:2007fw} or, equivalently, as two copies of the Nappi--Witten algebra \cite{Hartong:2017bwq,Penafiel:2019czp,Concha:2019lhn}. Indeed, dividing out \eqref{2xchiralalgebra} by the ideal generated by $\left\{  B^\pm_a \,, \; Z^\pm\,\right\}$ leads to the commutation relations
\begin{equation}\label{NWalgebra}
[ J^\pm,  G^\pm_{a}]=-\epsilon_{a}^{\;\;b}\,   G^\pm_{b}\,,\hskip 2.5 truecm
[  G^\pm_{a},  G^\pm_{b}]=\epsilon_{ab}\,  S^\pm \,.
\end{equation}
Redefining the generators as
\begin{equation}
P^\pm_a=\ell G^\pm_a\,,\hskip 2.5 truecm S^\pm=\ell^2 Z^\pm\,,
\end{equation}
the commutation relations \eqref{NWalgebra} take the form
\begin{equation}\label{NWalgebra2}
[ J^\pm,  P^\pm_{a}]=-\epsilon_{a}^{\;\;b}\,   P^\pm_{b}\,,\hskip 2.5 truecm
[  P^\pm_{a},  P^\pm_{b}]=\epsilon_{ab}\,  Z^\pm \,.
\end{equation}
This algebra is the universal central extension of the Euclidean algebra $\mathbb E_2$ and defines the Nappi-Witten algebra \cite{Figueroa-OFarrill:1994liu,Figueroa-OFarrill:1999cmq}, which can also be identified with the Maxwell algebra in 1+1 dimensions\footnote{In the original definition \cite{Isler:1989hq,Cangemi:1992bj,Nappi:1993ie}, the Nappi-Witten algebra was constructed as a central extension of the Poincar\'e algebra in 1+1 dimensions, which requires changing the signature of the spatial metric in \eqref{NWalgebra2}. In that case, \eqref{NWalgebra2} is isomorphic to the Maxwell algebra in two space-time dimensions\cite{Schrader:1972zd}.}.
Yet there is another interpretation of the algebra \eqref{NWalgebra} that follows from \cite{Alvarez:2007ys,Alvarez:2007fw}. In fact, performing the redefinition:
\begin{equation}
\bar H^\pm= -\ell J^\pm \,,\hskip 1.5 truecm
\bar G^\pm=G^\pm_1 \,,\hskip 1.5 truecm
\bar P^\pm=\ell G^\pm_2 \,,\hskip 1.5 truecm
\bar M^\pm=\ell S^\pm \,,
\end{equation}
the algebra \eqref{NWalgebra} can be written as the Bargmann--Newton-Hooke algebra in 1+1 dimensions
\begin{equation}
\left[ \bar G^\pm ,\bar H^\pm , \right]= \bar P^\pm \,,\hskip 1.5 truecm
\left[ \bar P^\pm, \bar H^\pm, \right]= - \frac{1}{\ell^2}\bar G^\pm \,,\hskip 1.5 truecm
\left[ \bar G^\pm, \bar P^\pm  \right]= \bar M^\pm\,.
\end{equation}
Therefore, depending on what interpretation we adopt, the chiral algebra presented in \eqref{2xchiralalgebra} can either define an extension of the Nappi-Witten algebra, the Newton-Hooke algebra in two dimensions, or even the Maxwell algebra in two dimensions.

Consider now the action for coadjoint AdS$_3$ gravity \eqref{CSchiralR} plus the $\mathfrak u (1)^2$ action \eqref{u12action} with global factor $\gamma_3$. Going to the chiral basis in the $\mathfrak u (1)^2$ sector requires to define the Abelian fields
\begin{equation}
\chi_\pm=a_1\pm a_2 \,.
\end{equation}
This leads to the chiral relativistic action \eqref{CSchiralR} plus the Abelian CS action
\begin{equation}\label{chiralabelianaction}
\gamma_3 \int \chi_\pm d\chi_\pm \,,
\end{equation}
which follows from considering two CS theories with connections
\begin{equation}
A^\pm=\Omega_\pm^A \mathcal J^\pm_A + E_\pm^A \mathcal P^\pm_A+\chi_\pm \mathcal Z^\pm\,,
\end{equation}
where the generator $\mathcal Z^\pm$ has been introduced in \eqref{chiralcontraction}.

Defining the NR gauge fields
\begin{equation}
A^\pm=\omega_\pm J^\pm+\omega_\pm^a G^\pm_a
+ s_\pm \mathcal S^\pm+ b_\pm^a B^\pm_a+z_\pm Z^\pm\,,
\end{equation}
and using \eqref{relabelchiral}, the contraction \eqref{chiralcontraction} induces the following relation between relativistic and NR fields
\begin{multicols}{2}
\begin{subequations} \label{RgaugefieldstoNR}
\setlength{\abovedisplayskip}{-12pt}
\allowdisplaybreaks
\begin{align}
\Omega_{\pm}^{0}&=  \omega_{\pm}+\frac{1}{2\varepsilon^{2}}s_{\pm}-\frac{1}{2\varepsilon^{4}}z_{\pm}\,,\\[.1truecm]
\Omega_{\pm}^{a}&=  \frac{1}{\varepsilon}\omega_{\pm}^{a}-\frac{1}{\varepsilon^{3}}b_{\pm}^{a}\,,\\[.1truecm]
\chi_{\pm}&=  \omega_{\pm}-\frac{1}{2\varepsilon^{2}}s_{\pm}+\frac{1}{2\varepsilon^{4}}z_{\pm}\,,\\[.1truecm]
\frac{1}{\ell}E_{\pm}^{0}&=  -\frac{1}{\varepsilon^{2}}s_{\pm}\,,\\[.1truecm]
\frac{1}{\ell}E_{\pm}^{a}&=-  \frac{1}{2\varepsilon}\omega_{\pm}^{a}-\frac{1}{2\varepsilon^{3}}b_{\pm}^{a}\,.
\end{align}
\end{subequations}
\end{multicols}\noindent
Substituting these expressions into the chiral action formed by \eqref{CSchiralR}  and \eqref{chiralabelianaction} leads, up to boundary terms, to the following action:
\begin{equation}
\begin{aligned}
S[A^\pm]=\frac{\ell}{2}&\int\Bigg[(-\gamma_{2}+\gamma_{3})\omega_\pm d\omega_\pm+\frac{(-\gamma_{1}+\gamma_{2})}{\varepsilon^{2}}\omega_\pm^{a}F^\pm_{a}[G^\pm]+\frac{2\gamma_{1}-\gamma_{2}-\gamma_{3}}{2\varepsilon^{2}}\left(s_\pm d\omega_\pm+\omega_\pm ds_\pm\right)\\
&\hskip 1.5 truecm
-\frac{\gamma_{2}}{\varepsilon^{4}}
\left(\omega_\pm^{a}F^\pm_{a}[B^\pm]+b_\pm^{a}F^\pm_{a}[G^\pm]\right)
+\frac{\gamma_{2}+\gamma_{3}}{2\varepsilon^{4}}\left(\omega_\pm  dz_\pm+z_\pm d\omega_\pm\right)
\\
&\hskip 1.5 truecm
+\frac{4\gamma_{1}-\gamma_{2}+\gamma_{3}}{4\varepsilon^{4}}s_\pm ds_\pm+\frac{3(\gamma_{1}-\gamma_{2})}{2\varepsilon^{4}}\epsilon_{ab}s_\pm \omega_\pm^{a}\omega_\pm^{b} + O(\varepsilon^{-6})\Bigg]\,.
\end{aligned}
\end{equation}
where we have defined
\begin{equation}
R^{a\pm}[G^\pm]=d\omega^a_\pm+\epsilon^a_{\;\;b}\omega_\pm \omega^b_\pm\,,
\hskip 1.5 truecm
R^{a\pm}[B^\pm]=d b^a_\pm+\epsilon^a_{\;\;b} \left(\omega_\pm b^b_\pm +s_\pm \omega^b_\pm \right)\,.
\end{equation}

Now we will consider three different choices of the parameters that lead to NR chiral actions:

\vskip .3truecm

\noindent \textbf{Abelian CS action}
\vskip .1truecm
In the case $\gamma_2 \ne \gamma_3$, the limit $\varepsilon\rightarrow \infty$ applied to the relativistic chiral action \eqref{CSchiralR} leads to the following simple result:
\begin{equation}
S_{\mathfrak u (1)}^\pm=-\frac{\ell\kappa}{2}\int \omega_\pm d\omega^\pm\,,
\end{equation}
where in this case we have set $\kappa=\gamma_2-\gamma_3$. This corresponds to an Abelian CS theory. The action for Galilean gravity \eqref{Gaction} can be recovered from this action by using \eqref{chiraldecaction}, i.e.,
\begin{equation}
S_{\textrm{Galilei}}=S_{\mathfrak u (1)}^+ -S_{\mathfrak u (1)}^- \,,
\end{equation}
and the relation
\begin{equation}\label{NRchiralfields1}
\omega_\pm = \omega \pm\frac{1}{\ell} \tau \,,
\end{equation}
which can be deduced from the change of basis \eqref{nonCintermsofC}. This decomposition unveils the minimal   symmetry of Galilei gravity \eqref{Gaction}, given by the $\mathfrak u(1)^2$ algebra.
\vskip .3truecm

\noindent \textbf{Nappi--Witten CS action}
\vskip .1truecm

The choice  $\gamma_1=0$ and $\gamma_2=\gamma_3=\varepsilon^2\kappa$ leads to the following NR chiral action
\begin{equation}\label{NWCS}
S_{\textrm{NW}}^\pm=\frac{\ell\kappa}{2} \int\left[\omega_{\pm}^{a}R_{a}^{\pm}[G^\pm]-\omega_\pm d s_\pm-s_\pm d \omega_\pm \right]\,.
\end{equation}
This CS action is invariant under the Nappi--Witten algebra \eqref{NWalgebra} and  has been studied from different points of views in \cite{Hartong:2017bwq,Penafiel:2019czp,Concha:2019lhn}.
The action for Extended Bargmann--Newton--Hooke gravity \eqref{Gaction} follows from the relation \eqref{chiraldecaction}, which in this case has the form
\begin{equation}
S_{\textrm{Extended-BNH}}=S_{\textrm{NW}}^+ -S_{\textrm{NW}}^-\,,
\end{equation}
and using the expression for the chiral fields \eqref{NRchiralfields1} and
\begin{equation}\label{NRchiralfields2}
s_\pm = s \pm\frac{1}{\ell} m \,,\hskip 2.5 truecm
\omega^a_\pm = \omega^a \pm\frac{1}{\ell} e^a\,.
\end{equation}
We should note that the action \eqref{NWCS} can also be obtained by setting $\gamma_2=\gamma_3=0$, together with the rescaling $\gamma_1=-\varepsilon^2\kappa$, in equation \eqref{CSchiralR}. In particular, this shows that it is possible to define a finite NR limit of the Einstein-Hilbert action without the addition of Abelian fields leading to a CS action invariant under the Nappi--Witten algebra.

\vskip .3truecm

\noindent \textbf{Enhanced Nappi--Witten CS action}
\vskip .1truecm

Choosing the parameters as $\gamma_1=\gamma_2=\gamma_3=-\varepsilon^4\kappa$ and taking the limit $\varepsilon\rightarrow\infty$ yields the chiral action
\begin{equation}\label{chiralENW}
S_{\textrm{ENW}} =\frac{\ell\kappa}{2}\int\left[\omega_{\pm}^{a}R_{a}^{\pm}[B]+b_{\pm}^{a}R_{a}^{\pm}[G]
-\omega_\pm d z_\pm-s_\pm d s_\pm-z_\pm d\omega_\pm\right] \,,
\end{equation}
which is invariant under the Enhancement of the Nappi--Witten algebra defined by \eqref{2xchiralalgebra}, and can be shown to define a CS action for that symmetry. Using the general result \eqref{chiraldecaction} and the relations \eqref{NRchiralfields1}, \eqref{NRchiralfields2} plus
\begin{equation}
z_\pm = z \pm\frac{1}{\ell} y \,,\hskip 2.5 truecm
b^a_\pm = b^a \pm\frac{1}{\ell} t^a \,,
\end{equation}
we find in this case
\begin{equation}\label{ENWaction}
S_{\textrm{Enhanced-BNH}}=S_{\textrm{ENW}}^+ -S_{\textrm{ENW}}^-\,.
\end{equation}
Thus, the chiral actions \eqref{ENWaction} can be put together to obtain the Enhanced Bargmann--Newton--Hooke gravity action \eqref{EBNHgrav}.

It is important to remark that the coadjoint AdS$_3$ algebra admits a second invariant tensor given by\footnote{This expression can be recognized as the coadjoint version of the exotic invariant bilinear form on $\mathfrak{so}(2,2)$ given by \eqref{exoticITads}. In the AdS$_3$ case, the corresponding CS action leads to the  exotic variant of three-dimensional gravity studied in \cite{Witten:1988hc}.}
\begin{equation}\label{invtensorcoadAdSex}
\left\langle  \tilde J_A   \tilde S_B  \right\rangle= \sigma_1 \eta_{AB} \,,\quad
\left\langle  \tilde P_A  \tilde  T_B  \right\rangle= \frac{\sigma_1}{\ell^2}\eta_{AB} \,,\quad
\left\langle  \tilde J_A   \tilde J_B  \right\rangle=  \sigma_2 \eta_{AB} \,,
\left\langle  \tilde P_A   \tilde P_B  \right\rangle=  \frac{\sigma_2}{\ell^2} \eta_{AB} \,,
\end{equation}
which is non-degenerate for $\sigma_1\neq0$. This can be used to define an exotic CS action for three-dimensional coadjoint AdS gravity. In the chiral basis this action can be obtained by considering
\begin{equation}
S_{CS}[A]=S[A^+]+S[A^-]
\end{equation}
instead of \eqref{chiraldecaction}. A NR limit of this theory can be constructed along the same lines that have been shown above. These NR actions in the standard (non-chiral) basis have been obtained by means of the Lie algebra expansion method in \cite{Gomis:2019nih}.

\section{ Conclusions and Generalization}

In this work we investigated in which sense some of the NR gravity actions that have appeared in the recent literature could be obtained as the NR limit of a relativistic action with an enhanced Poincar\'e symmetry. We focused on three-dimensional actions only. To describe these enhanced Poincar\'e symmetries a key role was played by the coadjoint Poincar\'e algebra. Specifically, we found that for specific choices of the parameters, the CS action based on the coadjoint Poincar\'e algebra has two finite NR limits: one leads to Galilei gravity and the other one to Extended Bargmann gravity. On the other hand, we showed that the CS action based on the coadjoint Poincar\'e $\ \oplus\ \mathfrak{u}(1)^2$ algebra has three NR limits determined by different choices of the parameters: Galilei gravity, Extended Bargmann gravity and a third new limit that yields the NR gravity action of \cite{Ozdemir:2019orp} which we denominated as Enhanced Bargmann gravity.

We were able to reproduce the NR algebra underlying the construction of \cite{Hansen:2018ofj,Hansen:2019vqf} by a particular contraction of the coadjoint Poincar\'e algebra. However, we could not find a NR limit of a three-dimensional coadjoint Poincar\'e invariant gravity action that leads to a NR gravity theory based on the algebra of  \cite{Hansen:2018ofj,Hansen:2019vqf}. Moreover, in \cite{Hansen:2018ofj} it was observed that the same NR algebra could be obtained from a contraction of the direct sum of the Poincar\'e and Euclidean algebras. It would be interesting to see if there is a relationship between this direct sum and the coadjoint Poincar\'e algebra we have been considering in this work.

It is natural to speculate about whether the results found in this work are part of a more general construction. At several places in our work we mentioned that the relevant algebra underlying our constructions could be obtained by a suitable quotient from the infinite-dimensional algebra \eqref{eq:GG} given in Appendix A.  One could ask oneself what are the relativistic counterparts of more general quotients of this infinite-dimensional algebra. Before doing that, it is instructive to summarize the pattern we found so far in this paper. We observe that each time a $\mathfrak{u}(1)^2$ factor is added to the previous algebra a new NR gravity action appears. To see how this goes, we start with the Poincar\'e algebra that produces the Galilei gravity action.\,\footnote{We note that the Galilei gravity action itself consists of two $\mathfrak{u}(1)$ gauge fields realizing a $\mathfrak{u}(1)^2$ algebra.} Extending to the Poincar\'e \ $\oplus\ \mathfrak{u}(1)^2$ algebra leads to the previous result of the Galilei gravity action plus the new Extended Bargmann gravity action. In a next step, extending to the coadjoint Poincar\'e algebra, reproduces the two actions we already had constructed. In contrast, a further extension to the coadjoint Poincar\'e \ $\oplus \ \mathfrak{u}(1)^2$ algebra leads to the previous result plus the new Enhanced Bargmann gravity action. Summarizing we have
\begin{eqnarray}
&&\textrm{Poincar\'e }\,:\hskip 3truecm \textrm{Galilei gravity}\,,\nonumber\\[.2truecm]
&&\textrm{Poincar\'e} \oplus \ \mathfrak{u}(1)^2 \,: \hskip 1.7truecm \textrm{previous plus Extended Bargmann gravity}\,, \label{pattern}\\[.2truecm]
&&\textrm{coadjoint Poincar\'e }\,:  \hskip 1.4truecm \textrm{previous}\,,\nonumber\\[.2truecm]
&&\textrm{coadjoint Poincar\'e} \oplus\ \mathfrak{u}(1)^2  :\hskip .1truecm \textrm{previous plus Enhanced Bargmann gravity\,}\nonumber
\end{eqnarray}

It is interesting to speculate about how the pattern \eqref{pattern} extends to larger algebras. Indeed, as shown in \cite{Bergshoeff:2019ctr,Gomis:2019nih,Gomis:2019fdh}, the infinite-dimensional algebra \eqref{eq:GG} can be obtained by considering a sequence of expansions of the Poincar\'e algebra.

In order to generate relativistic algebras beyond the coadjoint Poincar\'e algebra, it is instructive to note that the Poincar\'e algebra itself can be written as an expansion of the form
\begin{equation}
S_{E}^{(N)}\times \mathfrak{iso}(D-1,1)
\end{equation}
with $N=0$, while the coadjoint Poincar\'e algebra  corresponds the $N=1$ case\footnote{The $S_{E}^{(N)}$ semigroup has been introduced in \cite{Izaurieta:2006zz} to define expansions of Lie algebras.}.
Thus, an expansion of this form for $N=2$ case would triplicate the number of generators of the Poincar\'e algebra and is the natural candidate to define a relativistic counterpart of a larger truncation of the infinite-dimensional algebra \eqref{eq:GG}.

Regarding three-dimensional gravity actions, note that this mechanism does not provide a relativistic counterpart of the truncations of \eqref{eq:GG} that contain the required central extensions leading to well-defined NR limits. However, as it happens in the particular cases explored in this article, for $D=2+1$, these algebras can be conjectured to follow from contractions of direct products of the form
\begin{equation}
\left\{S_{E}^{(N)}\times \mathfrak{iso}(2,1)\right\}\oplus\mathfrak{u}(1)^2 \,.
\end{equation}
Checking the validity of the scenario sketched above would be a natural continuation of the results presented in this paper.

Finally, given the recent interest in the asymptotic symmetries of gravitational theories, we cannot resist to comment about the boundary dynamics of the NR gravities discussed in this paper. At the relativistic level, it is well-known that the asymptotic symmetry of three-dimensional AdS$_3$ gravity is given by two copies of the Virasoro algebra \cite{Brown:1986nw}. In the same way, given the fact that the coadjoint AdS$_3$ algebra is isomorphic to two copies of the Poincar\'e algebra and that the asymptotic symmetry of three-dimensional flat gravity is given by the $\mathfrak{bms}_3$ algebra \cite{Barnich:2006av}, it should be possible to find suitable boundary conditions for the gauge fields, such that the asymptotic symmetry of three-dimensional coadjoint AdS$_3$ gravity is given by the $\mathfrak{bms}_3 \oplus \mathfrak{bms}_3$ algebra. Thus, it would be interesting to investigate the fate of these relativistic asymptotic symmetries when considering NR limits. We hope to address these interesting issues in a future publication.

\section*{Acknowledgements}
We acknowledge discussions with Andrea Barducci, Roberto Casalbuoni, Axel Kleinschmidt, J{\o}rgen Sand{\o}e Musaeus, Niels Obers, Gerben Oling and Jakob Palmkvist. E.B. wishes to thank hospitality of Universitat de Barcelona where part of this work was done. J.G. and P.S-R. acknowledge the hospitality
and support of the Van Swinderen Institute where part of this work has been done. J.G. also has been supported in part by MINECO FPA2016-76005-C2-1-P and Consolider CPAN, and by the Spanish government (MINECO/FEDER) under project MDM-2014-0369 of ICCUB (Unidad de Excelencia Mar\`a de Maeztu). P.S-R. acknowledges DI-VRIEA for financial support through Proyecto Postdoctorado 2019 VRIEA-PUCV.

\appendix

\section{Infinite-dimensional algebra}
\label{appendixA}
Our starting point is the infinite-dimensional NR expansion of the Poincar\'e algebra  given by~\cite{Hansen:2019vqf,Gomis:2019fdh}
\begin{multicols}{2}
\begin{subequations} \label{eq:GG}
\setlength{\abovedisplayskip}{-13pt}
\allowdisplaybreaks
\begin{align}
[ B_a^{(m)},  H^{(n)},  ] &= P_a^{(m+n)}\,,\\[.1truecm]
[ B_a^{(m)} , P_b^{(n)} ] &= \delta_{ab} H^{(m+n+1)}\,,\\[.1truecm]
[ B_a^{(m)}, B_b^{(n)} ] &= J_{ab}^{(m+n+1)}\,,\\[.1truecm]
[J_{ab}^{(m)}, J_{cd}^{(n)}  ] &= 4\delta_{[c[b} J_{a]d]}^{(m+n)}  \,,\\[.1truecm]
[ J_{ab}^{(m)}, B_c^{(n)} ] &= 2\delta_{c[b} B_{a]}^{(m+n)}\,, \\[.1truecm]
[ J_{ab}^{(m)}, P_c^{(n)} ] &= 2\delta_{c[b} P_{a]}^{(m+n)}\,.
\end{align}
\end{subequations}
\end{multicols}\noindent
The stringy generalization of this kind of infinite algebra has been  considered in \cite{Harmark:2019upf,Bergshoeff:2019pij}, while the (A)dS extension has been found in \cite{Gomis:2019nih}.

The commutation relations of the NR limit of the coadjoint Poincar\'e algebra \eqref{contractioncoadp} can be viewed as the quotient of this infinite-dimensional algebra by the ideal generated by
\begin{equation}
\{ J_{ab}^{(m\geq 2)}, \; H^{(m\geq 2)},\;  B_a^{(n\geq 2)},\;  P_b^{(n\geq 2)}\}\,.
\end{equation}
Furthermore, considering the quotient of the infinite-dimensional algebra \eqref{eq:GG} by the smaller ideal
\begin{equation}
\{ J_{ab}^{(m\geq 2)}, \; H^{(m\geq 2)},\;  B_a^{(n\geq 1)},\;  P_b^{(n\geq 1)}\}\,,
\end{equation}
leads to a $D>3$ version of the Extended Bargmann algebra
\begin{multicols}{2}
\begin{subequations} \label{higherDExtBargmann}
\setlength{\abovedisplayskip}{-15pt}
\allowdisplaybreaks
\begin{align}
[   G_{a},   H]&=   P_{a}\, ,\\[.1truecm]
[   G_{a},   P_{b}]&=  \delta_{ab} M \, , \\[.1truecm]
[   G_{a},   G_{b}]&= S_{ab} \,,\\[.1truecm]
[   J_{ab},   J_{cd}]&=4\delta_{[a[c}   J_{d]b]}\, ,\\[.1truecm]
[   J_{ab},   S_{cd}]&=4\delta_{[a[c}   S_{d]b]}\, ,\\[.1truecm]
[   J_{ab},   P_{c}]&= 2\delta_{c[b}   P_{a]}\,,\\[.1truecm]
[   J_{ab},   G_{c}]&= 2\delta_{c[b}   G_{a]}\,.
\end{align}
\end{subequations}
\end{multicols}\noindent
In the same way, a $D>3$ version of the Enhanced Bargmann algebra defined by the commutation relations \eqref{Galilei}, \eqref{postNl2+1} and \eqref{extYZ3D} can be obtained considering the quotient by the ideal
\begin{equation}
\{ J_{ab}^{(m\geq 3)}, \; H^{(m\geq 3)},\;  B_a^{(n\geq 2)},\;  P_b^{(n\geq 2)}\}\,,
\end{equation}
which enlarges the algebra \eqref{higherDExtBargmann} to include the level two generators $H^{(2)}$ and $J_{ab}^{(2)}$.
Thus, we see that the same infinite-dimensional algebra can be used to generate different finite dimensional NR symmetries by considering quotients by suitable ideals.

\end{document}